# *Dual Flow*: Joint Image–Measurement Interpolation for Inverse Problems

Sevket Ege Yesilyurt and Figen S. Oktem, *Member, IEEE*

***Abstract*—Generative model-based approaches to linear inverse problems typically use the acquired noisy measurement to condition or guide sampling in the image domain. In this work, we introduce Dual Flow, a novel joint-flow model for solving linear inverse problems by jointly estimating the clean image and corresponding noise-free measurement from the observation. Specifically, Dual Flow formulates image reconstruction and measurement denoising as a coupled estimation problem, using coupled flows in the image and measurement domains. The acquired noisy measurement is used only to initialize the measurement state; thereafter, image and measurement states are jointly updated based on their current values. By construction, the final image and measurement estimates are guaranteed to satisfy the underlying forward model. We derive the sampling updates for the joint-flow which involves the conditional mean of the clean image given the joint state. We further derive a closed-form expression for a sufficient statistic of the joint state and show that the required conditional mean can be obtained by solving an image-domain Gaussian denoising problem involving operator-dependent, correlated noise. We train a velocity-prediction network using this formulation to estimate the conditional mean from the derived sufficient statistic. Extensive experiments across various linear inverse problems and image datasets demonstrate state-of-the-art performance in both distortion metrics and perceptual quality. Dual Flow requires no hyperparameter tuning beyond the number of inference steps and achieves shorter reconstruction times than competing methods. Additional analyses assess robustness to model mismatch, as well as the ability to generate diverse plausible reconstructions and characterize reconstruction uncertainty. Our code is available at https://github.com/METU-SPACE-Lab/Dual-Flow/.**

***Index Terms*—Inverse problems, image restoration, flow matching, diffusion models.**

## I. Introduction

INVERSE problems play a fundamental role in modern imaging and sensing, underpinning applications such as medical imaging [1], computational microscopy [2], remote sensing [3], as well as image restoration tasks such as super-resolution, deblurring, and inpainting [4]. In the context of linear inverse problems, the objective is to recover an unknown signal or image $\mathbf{x}_1 \in \mathbb{R}^n$ from noisy measurements $\mathbf{y}_0 \in \mathbb{R}^m$, which are often modeled as

$$\mathbf{y}_0 = \mathbf{A}\mathbf{x}_1 + \mathbf{n}. \tag{1}$$

Here the matrix $\mathbf{A} \in \mathbb{R}^{m\times n}$ represents the known forward (system) operator, and the noise vector $\mathbf{n}$ represents zero-mean white Gaussian noise.

Recovering the image $\mathbf{x}_1$ from the measurement $\mathbf{y}_0$ is often challenging, as the underlying system of linear equations may be underdetermined or ill-conditioned, leading to non-unique solutions or sensitivity to measurement noise. To address these challenges, reconstruction methods typically incorporate prior information about the unknown image to constrain the solution space and obtain reliable reconstructions. This prior information can be characterized by the prior density $p_{\widetilde{\mathbf{x}}_1}(\mathbf{x}_1)$ of the image, where $\widetilde{\mathbf{x}}_1$ denotes the corresponding clean image random variable. This prior distribution is generally unknown in practice.

A classical approach to incorporating prior information is through regularization-based reconstruction. From a Bayesian perspective, maximum posterior (MAP) estimation under the observation model in Eq. (1) leads to an optimization problem that involves a data-fidelity term and a regularization term corresponding to the negative log-prior (i.e., $-\log p_{\widetilde{\mathbf{x}}_1}(\mathbf{x}_1)$). Traditional regularization choices, such as total variation and sparsity-based regularization, typically promote smoothness or sparsity of the unknown image [5]–[7]. While the resulting optimization problems can often be efficiently solved using proximal splitting methods [8], [9], their reconstruction performance is often limited by the inability of hand-crafted regularizers to adequately capture complex image statistics.

To overcome the limitations of hand-crafted regularizers, deep learning has emerged as a powerful tool for developing data-driven image reconstruction methods. In plug-and-play (PnP) methods, the proximal operator, which can be interpreted as a MAP denoiser under an assumed image prior, is replaced by a learned denoiser, typically trained to minimize the mean squared error (MSE) [10]–[12]. Unrolled methods, on the other hand, convert a fixed number of iterations of a model-based reconstruction algorithm into a trainable network, combining learned image updates with explicit data-consistency steps that involve the forward operator. The learned parameters are typically specialized to the imaging setting used during training [13]–[15].

Despite these advances, many existing reconstruction methods aim to produce a single estimate of the unknown image from the given measurements, rather than generating diverse reconstructions that can enable the exploration of multiple plausible solutions and the characterization of reconstruction uncertainty, while potentially improving perceptual quality [16]–[18]. For instance, minimum mean-squared error (MMSE)-based reconstruction methods aim to approximate the posterior mean, which minimizes the expected squared reconstruction error but may lead to overly smooth images when the posterior distribution contains multiple distinct plausible solutions. MAP-based methods, on the other hand, seek to maximize the posterior density $p(\mathbf{x}_1|\mathbf{y}_0)$, yielding a single reconstruction without capturing the uncertainty associated with the inverse problem.

Generative models provide a powerful framework for learning complex image distributions from data and generating diverse plausible samples. In particular, diffusion models and flow matching have emerged as effective approaches for modeling high-dimensional image distributions [19]–[21]. Diffusion models generate new samples by reversing a gradual noise-corruption process, typically using learned score functions. Flow matching models, on the other hand, learn a time-dependent velocity field that transports samples from a simple source distribution to the target data distribution, often using straight-line interpolations between source and target samples during training [21], [22]. Both approaches can thus provide learned priors for image reconstruction. However, solving inverse problems requires incorporating the observed measurements into the generative process, rather than solely sampling from the learned image prior.

A major class of approaches for inverse problems addresses this by relying on pretrained unconditional generative models and incorporating measurement information only during inference. While these methods allow pretrained models to be reused across different inverse problems, accurately incorporating measurement information into the sampling process appears to be challenging. Many existing methods address this challenge through approximations of the likelihood score at intermediate stages to achieve posterior sampling [4], [23], [24], data-consistency corrections applied to the sampling process of a pretrained unconditional model [25]–[27], or Bayesian filtering techniques that employ weighted particles to approximate posterior sampling [28]. Despite their differences, these methods face limitations in posterior approximation accuracy and computational efficiency, as approximate measurement-conditioning strategies can lead to deviations from the true posterior [29], while Bayesian filtering methods can improve posterior approximation accuracy at the expense of increased computational cost.

While the aforementioned methods incorporate measurement information only during inference, an alternative class of approaches uses measurement information during training. For instance, conditional diffusion and flow matching methods train denoising or velocity-prediction networks using the acquired measurements or measurement-related quantities (such as $\mathbf{A}^\top\mathbf{y}_0$) as conditioning inputs [30]–[32]. Once trained, these methods can generate measurement-conditioned reconstructions without separate correction steps during inference. However, conditioning on measurements alone does not guarantee consistency with the underlying forward model (i.e. physics-based data consistency). Another line of work focuses on image restoration by learning deterministic or stochastic processes that progressively transform degraded observations into clean images [33]–[36]. More recently, posterior-score estimation has been reformulated as a denoising problem involving operator-dependent, correlated Gaussian noise, enabling a learned denoiser to estimate the posterior score directly and perform posterior sampling without approximations and separate data-consistency correction steps [37].

Existing generative model-based approaches to inverse problems typically use the acquired measurement $\mathbf{y}_0$ to condition or guide the sampling process in the image domain, rather than jointly modeling the image and the corresponding measurement. In this work, we introduce *Dual Flow*, a novel joint flow model over the image and measurement domains for solving linear inverse problems, which enables the joint estimation of the unknown image and the corresponding noise-free measurement. Specifically, under the forward model in Eq. (1), we define a joint image–measurement interpolation in which both the image and measurement states are defined through coupled straight-line interpolations. In particular, the image state is interpolated between an initial random sample and the clean image $\mathbf{x}_1$, while the measurement state is interpolated between the acquired noisy measurement $\mathbf{y}_0$ and the noise-free measurement $\mathbf{A}\mathbf{x}_1$ of the final clean image. Thus, the two coupled flows evolve jointly, with the image-domain flow mapping the initial noise process to a clean image estimate and the measurement-domain flow denoising the acquired measurement. These coupled interpolations provide the basis for learning joint flow dynamics in the image and measurement domains.

In our development, we combine the image and measurement states into a single variable, referred to as the joint state, and derive the sampling updates of this joint state for the reverse sampling procedure of Dual Flow. We show that these updates are governed by the conditional mean of the clean image random variable $\widetilde{\mathbf{x}}_1$ given the current joint state. To this end, we derive a closed-form sufficient statistic of the current joint state and show that estimating this conditional mean can be formulated as a denoising problem in which the clean image random variable $\widetilde{\mathbf{x}}_1$ is corrupted by operator-dependent correlated Gaussian noise. The derived sufficient statistic serves as the input to a velocity-prediction network, which is trained to obtain the conditional mean required for the joint-state sampling updates.

We conduct comprehensive experiments across different linear inverse problems and image datasets, and compare the reconstruction quality, perceptual quality, and computational efficiency of Dual Flow against recent approaches. We also perform additional analyses to investigate the effect of the number of inference steps, robustness to varying degradation levels and model mismatch, and the ability to generate multiple plausible reconstructions and characterize reconstruction uncertainty. Numerical results demonstrate that Dual Flow achieves state-of-the-art performance, outperforming competing methods in terms of both distortion metrics and perceptual quality, while requiring substantially less reconstruction time and no additional hyperparameters beyond the number of inference steps. Visual comparisons further suggest that Dual Flow better avoids over-smoothing and artifacts while preserving fine textures and sharp details. Overall, these results demonstrate that Dual Flow achieves substantial improvements in perceptual quality without sacrificing distortion performance, yielding a more favorable perception–distortion trade-off than competing methods.

Unlike existing approaches that use the acquired noisy measurement $\mathbf{y}_0$ to guide the reconstruction of the unknown image, Dual Flow formulates linear inverse problems as a joint estimation problem involving two coupled quantities: the clean image and its corresponding noise-free measurement. Specifically, it jointly addresses image reconstruction and measurement denoising within a single reverse sampling procedure, while ensuring that the two estimates are related through the forward operator. Rather than repeatedly using the fixed noisy measurement $\mathbf{y}_0$ to guide the sampling process, Dual Flow uses it only to initialize the measurement state, with subsequent updates relying on the evolving joint state. To enable this joint sampling procedure, we derive a closed-form expression for a sufficient statistic of the current joint state, which serves as the input to the velocity-prediction network for estimating the conditional mean of the clean image. This exact formulation allows the network to fully incorporate the relevant image and measurement information without relying on hand-crafted measurement-related quantities or approximate measurement-conditioning strategies. Furthermore, Dual Flow requires no hyperparameter tuning beyond selecting the number of inference steps, whereas existing approaches often require multiple hyperparameters to be tuned for different inverse problems.

The main contributions of this paper can be summarized as follows:

- We develop Dual Flow, a novel joint flow model for linear inverse problems that formulates image reconstruction and measurement denoising as a coupled estimation problem. The proposed model jointly estimates the clean image and its corresponding noise-free measurement through coupled straight-line interpolations in the image and measurement domains, with the two estimates directly related by the underlying forward operator.
- We derive the sampling updates for the joint state and show that they are governed by the conditional mean of the clean image given the current joint state. We further derive a closed-form sufficient statistic of the joint state and show that estimating this conditional mean can be reformulated as a denoising problem involving operator-dependent, correlated Gaussian noise.
- Based on this reformulation, we develop a velocity-prediction network that takes the derived sufficient statistic as input to estimate the conditional mean required for joint state sampling. By using the acquired noisy measurement only to initialize the measurement state and relying on the evolving joint state for subsequent updates, Dual Flow avoids the need for hand-crafted measurement-related conditioning inputs and approximate measurement-conditioning strategies. Moreover, the proposed method does not require hyperparameter tuning beyond selecting the number of inference steps.
- We conduct comprehensive experiments across various linear inverse problems and image datasets, demonstrating state-of-the-art performance in terms of both distortion metrics and perceptual quality, while requiring substantially less reconstruction time than competing methods. We also provide detailed analyses of the effects of the number of inference steps, robustness to varying degradation levels and model mismatch, reconstruction diversity, and uncertainty estimation.

The remainder of this paper is organized as follows. Section II provides the necessary background on conventional flow matching. Section III presents the Dual Flow method in detail, including its mathematical formulation and relevant derivations. Section IV presents numerical results for various linear inverse problems, compares Dual Flow against recent methods, and provides detailed analyses of its performance. Finally, Section V concludes the paper.

## II. Background

### A. Flow Matching

Flow matching learns a time-dependent velocity field that transports a source distribution $p_0$ to a data distribution $p_1$ with a continuous probability path indexed by time $t$ [21]. Let $\widetilde{\mathbf{x}}_0$ and $\widetilde{\mathbf{x}}_1$ denote random variables with the source and data distributions $p_0$ and $p_1$, respectively. A commonly used model for flow matching is based on straight-line interpolation [22]:

$$\widetilde{\mathbf{x}}_t = t\widetilde{\mathbf{x}}_1 + (1-t)\widetilde{\mathbf{x}}_0, \qquad t \in [0,1], \tag{2}$$

where $\widetilde{\mathbf{x}}_t$ denotes the intermediate random variable at time $t$ obtained through interpolation between the source random variable $\widetilde{\mathbf{x}}_0$ and the data random variable $\widetilde{\mathbf{x}}_1$, and $p_t$ denotes the probability distribution of $\widetilde{\mathbf{x}}_t$. Thus, $\{p_t\}_{t\in[0,1]}$ defines a path from the probability distribution $p_0$ (at $t=0$) to the distribution $p_1$ (at $t=1$). A time-dependent flow $f(t,\cdot)$ transports the source distribution $p_0$ to the data distribution $p_1$ along this probability path via the ordinary differential equation (ODE) [21], [22]

$$\frac{\partial}{\partial t} f(t, \mathbf{x}) = \mathbf{v}_t(f(t, \mathbf{x})), \qquad f(0, \mathbf{x}) = \mathbf{x}, \tag{3}$$

where $f(t, \mathbf{x})$ denotes the position at time $t$ of the ODE trajectory initialized at $\mathbf{x}$, and $\mathbf{v}_t$ denotes the marginal velocity field associated with this path. For the straight-line interpolation in Eq. (2), the resulting marginal velocity field is given by [22], [26]:

$$\mathbf{v}_t(\mathbf{x}_t) = \mathbb{E}[\widetilde{\mathbf{x}}_1 - \widetilde{\mathbf{x}}_0 \mid \widetilde{\mathbf{x}}_t = \mathbf{x}_t] \tag{4}$$

$$= \frac{\mathbb{E}[\widetilde{\mathbf{x}}_1 \mid \widetilde{\mathbf{x}}_t = \mathbf{x}_t] - \mathbf{x}_t}{1 - t}, \qquad t \in [0, 1). \tag{5}$$

The objective of flow matching is to approximate the marginal velocity field $\mathbf{v}_t(\mathbf{x}_t)$ using a neural network $\mathbf{v}_\theta(\mathbf{x}_t, t)$ by minimizing the following flow matching loss [21]:

$$\mathcal{L}_{\text{FM}}(\theta) = \mathbb{E}_{t \sim \mathcal{U}[0,1], \mathbf{x}_t \sim p_t} \left[ \|\mathbf{v}_\theta(\mathbf{x}_t, t) - \mathbf{v}_t(\mathbf{x}_t)\|_2^2 \right]. \tag{6}$$

However, the marginal velocity field is generally unavailable in practice. Conditional flow matching (CFM) [21], [38] instead minimizes the following mean squared error (MSE) loss:

$$\mathcal{L}_{\text{CFM}}(\theta) = \mathbb{E}_{t \sim \mathcal{U}[0,1], (\mathbf{x}_0, \mathbf{x}_1) \sim \pi} \left[ \left\| \mathbf{v}_\theta(\mathbf{x}_t, t) - \frac{\mathbf{x}_1 - \mathbf{x}_t}{1 - t} \right\|_2^2 \right], \tag{7}$$

where $\mathbf{x}_t = t\mathbf{x}_1 + (1 - t)\mathbf{x}_0$. This loss differs from the flow matching loss only by an additive constant independent of $\theta$. Here $\pi \in \Pi(p_0, p_1)$ denotes a coupling over $(\widetilde{\mathbf{x}}_0, \widetilde{\mathbf{x}}_1)$, given by joint distributions with marginals $p_0$ and $p_1$. After training with CFM loss, a data sample can be generated by drawing an initial sample $\mathbf{x}_0$ from the distribution $p_0$, i.e., $\mathbf{x}_0 \sim p_0$, and integrating Eq. (3) from $t = 0$ to $t = 1$, with $\mathbf{v}_t$ replaced by the learned velocity field $\mathbf{v}_\theta(\cdot, t)$.

## III. Dual Flow Approach

### A. Dual Flow Model

In this subsection, we introduce the joint image–measurement flow model in which both the image and measurement states evolve jointly over time, rather than updating only the image state and using the acquired noisy measurement as fixed. We first present the main equations of the developed method, followed by their derivations and interpretations in the subsequent subsections.

Let $\widetilde{\mathbf{x}}_1$ denote the clean image random variable with distribution $p_{\text{data}}$ and $\widetilde{\mathbf{x}}_0$ denote the initial random variable with distribution $\mathcal{N}(\mathbf{0}, \mathbf{I})$ and independent of $\widetilde{\mathbf{x}}_1$. Similarly, let $\widetilde{\mathbf{y}}_1 = \mathbf{A}\widetilde{\mathbf{x}}_1$ denote the corresponding noise-free measurement and $\widetilde{\mathbf{y}}_0 = \mathbf{A}\widetilde{\mathbf{x}}_1 + \widetilde{\mathbf{n}}$ denote the noisy measurement, where the random variable $\widetilde{\mathbf{n}}$ denoting noise has distribution $\mathcal{N}(\mathbf{0}, \sigma^2\mathbf{I})$ and is independent of $(\widetilde{\mathbf{x}}_0, \widetilde{\mathbf{x}}_1)$. The straight-line interpolations in the image and measurement domains are defined as follows:

$$\widetilde{\mathbf{x}}_t = t\widetilde{\mathbf{x}}_1 + (1 - t)\widetilde{\mathbf{x}}_0, \tag{8}$$

$$\widetilde{\mathbf{y}}_t = t\widetilde{\mathbf{y}}_1 + (1 - t)\widetilde{\mathbf{y}}_0, \tag{9}$$

where $t \in [0, 1]$. We refer to $\widetilde{\mathbf{x}}_t$ and $\widetilde{\mathbf{y}}_t$ as the image- and measurement-domain states, respectively. Combining these states together, we define the joint state random variable $\widetilde{\mathbf{u}}_t = [(\widetilde{\mathbf{x}}_t)^T \ \ (\widetilde{\mathbf{y}}_t)^T]^T$. The joint image–measurement interpolation can then be expressed in terms of $\widetilde{\mathbf{u}}_t$ as follows:

$$\widetilde{\mathbf{u}}_t = t\widetilde{\mathbf{u}}_1 + (1 - t)\widetilde{\mathbf{u}}_0. \tag{10}$$

*Dual Flow Sampling:* Let $N$ denote the number of Euler steps, with step size $\Delta = 1/N$ and $t_k = k\Delta$ for $k = 0, \ldots, N$. Starting from $\mathbf{u}^{(0)} = [\mathbf{x}_0^T \ \ \mathbf{y}_0^T]^T$, which contains an initial random sample $\mathbf{x}_0 \sim \mathcal{N}(\mathbf{0}, \mathbf{I})$ and the acquired noisy measurement $\mathbf{y}_0$ (an observed realization of $\widetilde{\mathbf{y}}_0$), the sampling update for the joint-state sample takes the form

$$\mathbf{u}^{(k+1)} = \mathbf{u}^{(k)} + \frac{\Delta}{1 - t_k} \left( \begin{bmatrix} \mathbf{I} \\ \mathbf{A} \end{bmatrix} \mathbb{E}[\widetilde{\mathbf{x}}_1 \mid \widetilde{\mathbf{u}}_{t_k} = \mathbf{u}^{(k)}] - \mathbf{u}^{(k)} \right), \tag{11}$$

for $k = 0, \ldots, N - 1$, where $\mathbf{u}^{(k)} = [(\mathbf{x}^{(k)})^T \ \ (\mathbf{y}^{(k)})^T]^T$, with $\mathbf{x}^{(0)} = \mathbf{x}_0$ and $\mathbf{y}^{(0)} = \mathbf{y}_0$.

*Conditional Mean* $\mathbb{E}[\widetilde{\mathbf{x}}_1 \mid \widetilde{\mathbf{u}}_t = \mathbf{u}_t]$ *via Denoising*: The joint-state update requires the evaluation of the conditional mean $\mathbb{E}[\widetilde{\mathbf{x}}_1 \mid \widetilde{\mathbf{u}}_t = \mathbf{u}_t]$, where $\mathbf{u}_t = [\mathbf{x}_t^T \ \ \mathbf{y}_t^T]^T$ denotes a given value of $\widetilde{\mathbf{u}}_t$ at time $t$. For this, we utilize $\mathbf{q}_t$ defined as follows:

$$\mathbf{q}_t = \left(t^2\mathbf{I} + \sigma^{-2}\mathbf{A}^T\mathbf{A}\right)^{-1} \left[t\mathbf{I} \quad \sigma^{-2}\mathbf{A}^T\right] \mathbf{u}_t \tag{12}$$

Let $\widetilde{\mathbf{q}}_t$ denote the random variable obtained by applying the transformation in Eq. (12) to $\widetilde{\mathbf{u}}_t$. Here, $\widetilde{\mathbf{q}}_t$ provides a sufficient statistic for estimating $\widetilde{\mathbf{x}}_1$ from $\widetilde{\mathbf{u}}_t$, resulting in the following conditional mean equivalence:

$$\mathbb{E}[\widetilde{\mathbf{x}}_1 \mid \widetilde{\mathbf{u}}_t = \mathbf{u}_t] = \mathbb{E}[\widetilde{\mathbf{x}}_1 \mid \widetilde{\mathbf{q}}_t = \mathbf{q}_t]. \tag{13}$$

The defined $\widetilde{\mathbf{q}}_t$ can also be viewed as a noisy observation of $\widetilde{\mathbf{x}}_1$ with correlated Gaussian noise, hence estimating $\widetilde{\mathbf{x}}_1$ from the given $\mathbf{q}_t$ reduces to a denoising problem that can be effectively addressed using neural networks.

*Velocity-Prediction Network:* We then train a velocity-prediction network $V_\theta(t\mathbf{q}_t, \mathbf{y}_t, t)$ on this denoising problem. Then $\widehat{\mathbf{x}}_1(\mathbf{u}_t, t)$, representing the estimate of the conditional mean $\mathbb{E}[\widetilde{\mathbf{x}}_1 \mid \widetilde{\mathbf{u}}_t = \mathbf{u}_t]$, is obtained using this trained network as follows:

$$\widehat{\mathbf{x}}_1(\mathbf{u}_t, t) = t\mathbf{q}_t + (1 - t)V_\theta(t\mathbf{q}_t, \mathbf{y}_t, t). \tag{14}$$

Using this, the joint-state sampling update in Eq. (11) can be implemented. We next provide the derivations and details of these steps.

### B. Dual Flow Sampling

In this subsection, we derive the joint-state sampling update in Eq. (11) and discuss how the image- and measurement-domain state samples simultaneously evolve conditioned on the current joint-state sample.

Using the Euler method with number of inference steps $N$ and step size $\Delta = 1/N$, the joint state can be updated as

$$\mathbf{u}^{(k+1)} = \mathbf{u}^{(k)} + \Delta \mathbf{v}_{t_k}(\mathbf{u}^{(k)}). \tag{15}$$

Using Eq. (10) and the expression of the marginal velocity field given in Eq. (5), this update can be rewritten as

$$\begin{aligned}\mathbf{u}^{(k+1)} &= \mathbf{u}^{(k)} + \frac{\Delta}{1-t_k}\left(\mathbb{E}[\widetilde{\mathbf{u}}_1 \mid \widetilde{\mathbf{u}}_{t_k} = \mathbf{u}^{(k)}] - \mathbf{u}^{(k)}\right) \\ &= \mathbf{u}^{(k)} + \frac{\Delta}{1-t_k}\left(\begin{bmatrix}\mathbf{I}\\ \mathbf{A}\end{bmatrix}\mathbb{E}[\widetilde{\mathbf{x}}_1 \mid \widetilde{\mathbf{u}}_{t_k} = \mathbf{u}^{(k)}] - \mathbf{u}^{(k)}\right). \end{aligned} \tag{16}$$

Here, the last equality follows from $\widetilde{\mathbf{u}}_1 = [(\widetilde{\mathbf{x}}_1)^T \quad (\widetilde{\mathbf{y}}_1)^T]^T = [\mathbf{I} \quad \mathbf{A}^T]^T \widetilde{\mathbf{x}}_1$. Hence, we reach the joint-state update in Eq. (11), which involves $\mathbb{E}[\widetilde{\mathbf{x}}_1 \mid \widetilde{\mathbf{u}}_{t_k} = \mathbf{u}^{(k)}]$, the conditional mean of the clean image given the current joint-state sample. To explicitly formulate the evolution of the image- and measurement-domain state samples, equivalently, we can express $\mathbf{x}^{(k+1)}$ and $\mathbf{y}^{(k+1)}$ as follows:

$$\mathbf{x}^{(k+1)} = \mathbf{x}^{(k)} + \frac{\Delta}{1-t_k}\left(\mathbb{E}[\widetilde{\mathbf{x}}_1 \mid \widetilde{\mathbf{u}}_{t_k} = \mathbf{u}^{(k)}] - \mathbf{x}^{(k)}\right), \tag{17}$$

$$\mathbf{y}^{(k+1)} = \mathbf{y}^{(k)} + \frac{\Delta}{1-t_k}\left(\mathbf{A}\mathbb{E}[\widetilde{\mathbf{x}}_1 \mid \widetilde{\mathbf{u}}_{t_k} = \mathbf{u}^{(k)}] - \mathbf{y}^{(k)}\right). \tag{18}$$

The reverse sampling procedure starts at $t_0 = 0$ with the acquired noisy measurement $\mathbf{y}_0$ and an initial random sample $\mathbf{x}_0 \sim \mathcal{N}(\mathbf{0}, \mathbf{I})$, which together form the initial joint-state sample $\mathbf{u}^{(0)}$. To obtain the final estimate $\mathbf{u}^{(N)}$ at $t_N = 1$, the joint-state sample $\mathbf{u}^{(k)}$ is required to be iteratively updated using Eq. (11).

Having derived Eq. (11), we now discuss the behavior of this dual flow. Note that, using the expression of the marginal velocity field given in Eq. (4), $\mathbf{u}^{(k+1)}$ can also be expressed as

$$\begin{aligned}\mathbf{u}^{(k+1)} &= \mathbf{u}^{(k)} + \Delta\mathbb{E}[\widetilde{\mathbf{u}}_1 - \widetilde{\mathbf{u}}_0 \mid \widetilde{\mathbf{u}}_{t_k} = \mathbf{u}^{(k)}] \\ &= \mathbb{E}\big[(t_k + \Delta)\widetilde{\mathbf{u}}_1 + (1 - (t_k + \Delta))\widetilde{\mathbf{u}}_0 \mid \widetilde{\mathbf{u}}_{t_k} = \mathbf{u}^{(k)}\big] \\ &= \mathbb{E}[\widetilde{\mathbf{u}}_{t_{k+1}} \mid \widetilde{\mathbf{u}}_{t_k} = \mathbf{u}^{(k)}]. \end{aligned} \tag{19}$$

Or equivalently,

$$\mathbf{x}^{(k+1)} = \mathbb{E}\big[\widetilde{\mathbf{x}}_{t_{k+1}} \mid \widetilde{\mathbf{x}}_{t_k} = \mathbf{x}^{(k)}, \widetilde{\mathbf{y}}_{t_k} = \mathbf{y}^{(k)}\big], \tag{20}$$

$$\mathbf{y}^{(k+1)} = \mathbb{E}\big[\widetilde{\mathbf{y}}_{t_{k+1}} \mid \widetilde{\mathbf{x}}_{t_k} = \mathbf{x}^{(k)}, \widetilde{\mathbf{y}}_{t_k} = \mathbf{y}^{(k)}\big]. \tag{21}$$

Hence, the update of each individual state sample is based on its conditional expectation given the current joint-state sample $\mathbf{u}^{(k)}$.

Moreover, using Eqs. (8) and (9), the direct relationship between the intermediate states $\widetilde{\mathbf{x}}_t$ and $\widetilde{\mathbf{y}}_t$ can be expressed as

$$\widetilde{\mathbf{y}}_t = \mathbf{A}\widetilde{\mathbf{x}}_t + (1-t)(\widetilde{\mathbf{y}}_0 - \mathbf{A}\widetilde{\mathbf{x}}_0). \tag{22}$$

Notably, whereas the observation model in Eq. (1) relates the clean image $\mathbf{x}_1$ to the acquired noisy measurement $\mathbf{y}_0$ using the forward operator $\mathbf{A}$, Eq. (22) provides a similar relationship between the intermediate states $\widetilde{\mathbf{x}}_t$ and $\widetilde{\mathbf{y}}_t$. In particular, the discrepancy $\widetilde{\mathbf{y}}_t - \mathbf{A}\widetilde{\mathbf{x}}_t$ equals the initial residual $\widetilde{\mathbf{y}}_0 - \mathbf{A}\widetilde{\mathbf{x}}_0$, scaled by $(1-t)$. As $t$ increases, this discrepancy decreases and vanishes at $t = 1$, yielding $\widetilde{\mathbf{y}}_1 = \mathbf{A}\widetilde{\mathbf{x}}_1$.

The same relationship also holds for the image and measurement state samples during reverse sampling. To show this, let $\mathbf{B} = [-\mathbf{A} \quad \mathbf{I}]$, which satisfies $\mathbf{B}\widetilde{\mathbf{u}}_1 = \mathbf{0}$. Applying $\mathbf{B}$ to Eq. (16), we obtain

$$\mathbf{B}\mathbf{u}^{(k+1)} = \left(1 - \frac{\Delta}{1-t_k}\right)\mathbf{B}\mathbf{u}^{(k)}. \tag{23}$$

Since $1 - \frac{\Delta}{1-t_k} = \frac{1-t_{k+1}}{1-t_k}$ and $t_0 = 0$, starting from $k = 0$ and recursively applying Eq. (23), we obtain $\mathbf{B}\mathbf{u}^{(k)} = (1-t_k)\mathbf{B}\mathbf{u}^{(0)}$, or equivalently,

$$\mathbf{y}^{(k)} = \mathbf{A}\mathbf{x}^{(k)} + (1-t_k)(\mathbf{y}_0 - \mathbf{A}\mathbf{x}_0). \tag{24}$$

Thus, the same relationship as in Eq. (22) is preserved during sampling, yielding $\mathbf{y}^{(N)} = \mathbf{A}\mathbf{x}^{(N)}$ for the final estimate.

In summary, our sampling update in Eq. (11) jointly evolves the image- and measurement-domain state samples while maintaining the relationship between them in Eq. (24). In particular, the image state sample evolves from the initial random sample $\mathbf{x}_0$ toward the clean image estimate $\mathbf{x}^{(N)}$, while the measurement state sample evolves from the acquired noisy measurement $\mathbf{y}_0$ toward the noise-free measurement $\mathbf{y}^{(N)} = \mathbf{A}\mathbf{x}^{(N)}$ of the clean image estimate $\mathbf{x}^{(N)}$. Note that the acquired noisy measurement $\mathbf{y}_0$ is used as the measurement-domain state sample only in the initial step at $t_0 = 0$, whereas subsequent update steps use the evolving intermediate measurement state sample $\mathbf{y}^{(k)}$.

Because the joint-state sampling update requires the conditional mean $\mathbb{E}[\widetilde{\mathbf{x}}_1 \mid \widetilde{\mathbf{u}}_t = \mathbf{u}_t]$, we next establish the equivalent denoising formulation to estimate this conditional mean.

### C. Conditional Mean $\mathbb{E}[\widetilde{\mathbf{x}}_1 \mid \widetilde{\mathbf{u}}_t = \mathbf{u}_t]$ via Denoising

We now show that the conditional mean $\mathbb{E}[\widetilde{\mathbf{x}}_1 \mid \widetilde{\mathbf{u}}_t = \mathbf{u}_t]$ is equivalent to $\mathbb{E}[\widetilde{\mathbf{x}}_1 \mid \widetilde{\mathbf{q}}_t = \mathbf{q}_t]$, which can be obtained by solving a denoising problem on $\widetilde{\mathbf{x}}_1$ with correlated Gaussian noise and noisy observation $\mathbf{q}_t$ given in Eq. (12).

Let $\mathbf{H}_t = \begin{bmatrix} t\mathbf{I} & \mathbf{A}^T \end{bmatrix}^T$. The joint image–measurement interpolation can then be re-expressed as the following linear observation model:

$$\widetilde{\mathbf{u}}_t = \mathbf{H}_t \widetilde{\mathbf{x}}_1 + (1-t)\begin{bmatrix}\widetilde{\mathbf{x}}_0 \\ \widetilde{\mathbf{n}}\end{bmatrix}, \tag{25}$$

where the additive noise term is zero-mean Gaussian noise with covariance

$$\mathbf{R}_t = (1-t)^2 \begin{bmatrix}\mathbf{I} & \mathbf{0} \\ \mathbf{0} & \sigma^2\mathbf{I}\end{bmatrix}. \tag{26}$$

By Bayes' rule, the posterior $p_{\widetilde{\mathbf{x}}_1|\widetilde{\mathbf{u}}_t=\mathbf{u}_t}(\mathbf{x}_1)$ satisfies

$$p_{\widetilde{\mathbf{x}}_1|\widetilde{\mathbf{u}}_t=\mathbf{u}_t}(\mathbf{x}_1) = p_{\widetilde{\mathbf{u}}_t|\widetilde{\mathbf{x}}_1=\mathbf{x}_1}(\mathbf{u}_t)p_{\widetilde{\mathbf{x}}_1}(\mathbf{x}_1)/p_{\widetilde{\mathbf{u}}_t}(\mathbf{u}_t) \\ \propto \exp\left(-\frac{1}{2}\|\mathbf{u}_t - \mathbf{H}_t\mathbf{x}_1\|^2_{\mathbf{R}_t^{-1}}\right) p_{\widetilde{\mathbf{x}}_1}(\mathbf{x}_1). \tag{27}$$

where the terms that depend only on $\mathbf{u}_t$ are dropped. Here, $\|\cdot\|_{\mathbf{W}}$ denotes the weighted norm with a positive-definite matrix $\mathbf{W}$.

We then define the whitened variable $\bar{\mathbf{u}}_t = \mathbf{R}_t^{-1/2}\mathbf{u}_t$ and $\bar{\mathbf{H}}_t = \mathbf{R}_t^{-1/2}\mathbf{H}_t$. The weighted norm term in Eq. (27) can be expressed in terms of these variables as follows:

$$\begin{aligned} \|\mathbf{u}_t - \mathbf{H}_t\mathbf{x}_1\|^2_{\mathbf{R}_t^{-1}} &= \|\bar{\mathbf{u}}_t - \bar{\mathbf{H}}_t\mathbf{x}_1\|_2^2 \\ &= \left\|\left(\bar{\mathbf{H}}_t\bar{\mathbf{H}}_t^+\bar{\mathbf{u}}_t - \bar{\mathbf{H}}_t\mathbf{x}_1\right) + \left(\mathbf{I} - \bar{\mathbf{H}}_t\bar{\mathbf{H}}_t^+\right)\bar{\mathbf{u}}_t\right\|_2^2 \\ &= \left\|\bar{\mathbf{H}}_t\left(\bar{\mathbf{H}}_t^+\bar{\mathbf{u}}_t - \mathbf{x}_1\right)\right\|_2^2 + \left\|\left(\mathbf{I} - \bar{\mathbf{H}}_t\bar{\mathbf{H}}_t^+\right)\bar{\mathbf{u}}_t\right\|_2^2 \\ &= \left\|\bar{\mathbf{H}}_t^+\bar{\mathbf{u}}_t - \mathbf{x}_1\right\|^2_{\bar{\mathbf{H}}_t^T\bar{\mathbf{H}}_t} + \text{const.} \end{aligned} \tag{28}$$

where the third equality follows from decomposition into components in the range space of $\bar{\mathbf{H}}_t$ and its orthogonal complement since $\bar{\mathbf{H}}_t\bar{\mathbf{H}}_t^+$ is the orthogonal projector onto the range space of $\bar{\mathbf{H}}_t$ with $\bar{\mathbf{H}}_t^+$ denoting the Moore–Penrose pseudoinverse of $\bar{\mathbf{H}}_t$. The constant term in the final equality is a term that does not depend on $\mathbf{x}_1$.

We now define $\mathbf{q}_t = \bar{\mathbf{H}}_t^+\bar{\mathbf{u}}_t$, which can be expressed for $t \in (0,1)$ as follows:

$$\begin{aligned} \mathbf{q}_t &= \left(\bar{\mathbf{H}}_t^T\bar{\mathbf{H}}_t\right)^{-1}\bar{\mathbf{H}}_t^T\bar{\mathbf{u}}_t \\ &= \left(\mathbf{H}_t^T\mathbf{R}_t^{-1}\mathbf{H}_t\right)^{-1}\mathbf{H}_t^T\mathbf{R}_t^{-1}\mathbf{u}_t \\ &= \left(t^2\mathbf{I} + \sigma^{-2}\mathbf{A}^T\mathbf{A}\right)^{-1}\left[t\mathbf{I} \quad \sigma^{-2}\mathbf{A}^T\right]\mathbf{u}_t, \end{aligned} \tag{29}$$

where we use the fact that $\bar{\mathbf{H}}_t$ has full column rank for $t \in (0,1)$, hence $\bar{\mathbf{H}}_t^T\bar{\mathbf{H}}_t$ is invertible. This is the $\mathbf{q}_t$ expression given in Eq. (12).

Substituting Eq. (28) into Eq. (27) and using the defined variable $\mathbf{q}_t$, the posterior becomes

$$p_{\widetilde{\mathbf{x}}_1|\widetilde{\mathbf{u}}_t=\mathbf{u}_t}(\mathbf{x}_1) \propto \exp\left(-\frac{1}{2}\|\mathbf{q}_t - \mathbf{x}_1\|^2_{\bar{\mathbf{H}}_t^T\bar{\mathbf{H}}_t}\right)p_{\widetilde{\mathbf{x}}_1}(\mathbf{x}_1). \tag{30}$$

Under Bayesian estimation framework, the right-hand side of Eq. (30) is associated with the following observation model:

$$\widetilde{\mathbf{q}}_t = \widetilde{\mathbf{x}}_1 + \widetilde{\mathbf{n}}_t, \tag{31}$$

where $\widetilde{\mathbf{n}}_t$ is distributed with $\mathcal{N}\left(\mathbf{0}, (\bar{\mathbf{H}}_t^T\bar{\mathbf{H}}_t)^{-1}\right)$. This corresponds to a denoising problem in the image domain with correlated Gaussian noise. Moreover, the posterior $p_{\widetilde{\mathbf{x}}_1|\widetilde{\mathbf{q}}_t=\mathbf{q}_t}(\mathbf{x}_1)$ associated with the model in Eq. (31) is identical to the posterior $p_{\widetilde{\mathbf{x}}_1|\widetilde{\mathbf{u}}_t=\mathbf{u}_t}(\mathbf{x}_1)$ associated with the observation model given in Eq. (25). As a result,

$$\mathbb{E}[\widetilde{\mathbf{x}}_1 \mid \widetilde{\mathbf{u}}_t = \mathbf{u}_t] = \mathbb{E}[\widetilde{\mathbf{x}}_1 \mid \widetilde{\mathbf{q}}_t = \mathbf{q}_t]. \tag{32}$$

Hence, we obtain the conditional mean equivalence in Eq. (13).

Note that the noisy observation $\widetilde{\mathbf{q}}_t$ has the same dimension as the image state $\widetilde{\mathbf{x}}_t$, rather than the joint state $\widetilde{\mathbf{u}}_t = [(\widetilde{\mathbf{x}}_t)^T \ \ (\widetilde{\mathbf{y}}_t)^T]^T$, while retaining all the information in $\widetilde{\mathbf{u}}_t$ relevant to estimating the clean image $\widetilde{\mathbf{x}}_1$. Thus, $\widetilde{\mathbf{q}}_t$ is a sufficient statistic of $\widetilde{\mathbf{u}}_t$ for estimating $\widetilde{\mathbf{x}}_1$.

This reformulation and the success of neural networks in image denoising [19], [39] motivate us to estimate the conditional mean $\mathbb{E}[\widetilde{\mathbf{x}}_1 \mid \widetilde{\mathbf{u}}_t = \mathbf{u}_t]$ using the equivalent observation $\mathbf{q}_t$ through an image-domain Gaussian denoising problem, rather than directly using the joint state $\mathbf{u}_t$ under the original problem in Eq. (25). Based on this, we next describe how to train a velocity-prediction network for this conditional mean.

### *D. Velocity-Prediction Network*

In this subsection, we explain the training of a velocity-prediction network $V_\theta$ on this denoising problem to estimate the conditional mean $\mathbb{E}[\widetilde{\mathbf{x}}_1 \mid \widetilde{\mathbf{u}}_t = \mathbf{u}_t]$. This estimate is then used to perform Dual Flow sampling with Eq. (11).

We first scale the noisy observation $\widetilde{\mathbf{q}}_t$ in Eq. (31) by $t$, defining $\bar{\widetilde{\mathbf{q}}}_t = t\widetilde{\mathbf{q}}_t$, to match the $t\widetilde{\mathbf{x}}_1$ term in the image-domain interpolation formula given in Eq. (8). Let $\bar{\mathbf{q}}_t = t\mathbf{q}_t$ denote a given value of $\bar{\widetilde{\mathbf{q}}}_t$ at time $t$. We then define a velocity parametrization analogous to Eq. (5):

$$\bar{\mathbf{v}}_t(\bar{\mathbf{q}}_t) = \frac{\mathbb{E}[\widetilde{\mathbf{x}}_1 \mid \bar{\widetilde{\mathbf{q}}}_t = \bar{\mathbf{q}}_t] - \bar{\mathbf{q}}_t}{1-t}, \qquad t \in (0,1). \tag{33}$$

To learn this, we train a velocity-prediction network $V_\theta$ using the scaled equivalent observation $\bar{\mathbf{q}}_t$ and the measurement-domain state $\mathbf{y}_t$ with the following MSE loss:

$$\min_\theta \mathbb{E}_{t\sim\mathcal{U}(0,1),\mathbf{x}_1,\mathbf{x}_0,\mathbf{n}}\left[\left\|V_\theta(\bar{\mathbf{q}}_t, \mathbf{y}_t, t) - \frac{\mathbf{x}_1 - \bar{\mathbf{q}}_t}{1-t}\right\|_2^2\right]. \tag{34}$$

The network input $\bar{\mathbf{q}}_t$ in Eq. (34) can be computed as follows for $t \in (0,1)$:

$$\begin{aligned} \bar{\mathbf{q}}_t &= t\mathbf{q}_t \\ &= \left(t^2\mathbf{I} + \sigma^{-2}\mathbf{A}^T\mathbf{A}\right)^{-1}\left[t^2\mathbf{I} \quad t\sigma^{-2}\mathbf{A}^T\right]\mathbf{u}_t. \end{aligned} \tag{35}$$

During training, the joint state $\mathbf{u}_t$ is constructed based on the joint interpolation formula in Eq. (10) by independently drawing samples $\mathbf{x}_1 \sim p_{\text{data}}$, $\mathbf{x}_0 \sim \mathcal{N}(\mathbf{0}, \mathbf{I})$, and $\mathbf{n} \sim \mathcal{N}(\mathbf{0}, \sigma^2\mathbf{I})$, with $t \sim \mathcal{U}(0,1)$. The corresponding network input $\bar{\mathbf{q}}_t$ is then computed using Eq. (35). The resulting training procedure is summarized in Algorithm 1.

Having described the training procedure of Dual Flow in Algorithm 1, we now focus on the reverse sampling procedure. For a minimizer $V^\star$ of the MSE loss in Eq. (34), we have

$$\begin{aligned} V^\star(\bar{\mathbf{q}}_t, \mathbf{y}_t, t) &= \frac{\mathbb{E}[\widetilde{\mathbf{x}}_1 \mid \bar{\widetilde{\mathbf{q}}}_t = \bar{\mathbf{q}}_t, \widetilde{\mathbf{y}}_t = \mathbf{y}_t] - \bar{\mathbf{q}}_t}{1-t} \\ &= \frac{\mathbb{E}[\widetilde{\mathbf{x}}_1 \mid \widetilde{\mathbf{u}}_t = \mathbf{u}_t] - \bar{\mathbf{q}}_t}{1-t}, \end{aligned} \tag{36}$$

where the last equality follows directly from Eq. (32) for $t \in (0,1)$, since the given information is equivalent, for estimating $\widetilde{\mathbf{x}}_1$, to conditioning solely on $\widetilde{\bar{\mathbf{q}}}_t = \bar{\mathbf{q}}_t$. At $t = 0$, note that $\mathbb{E}[\widetilde{\mathbf{x}}_1 \mid \widetilde{\mathbf{u}}_0 = \mathbf{u}_0] = \mathbb{E}[\widetilde{\mathbf{x}}_1 \mid \widetilde{\mathbf{y}}_0 = \mathbf{y}_0]$ since $\widetilde{\mathbf{x}}_0$ is independent of $(\widetilde{\mathbf{x}}_1, \widetilde{\mathbf{y}}_0)$. Moreover, additionally conditioning on the initial value of $\widetilde{\bar{\mathbf{q}}}_t$, which is a function of $\widetilde{\mathbf{u}}_0$, does not affect this conditional mean.

Using Eq. (36), the conditional mean satisfies

$$\mathbb{E}[\widetilde{\mathbf{x}}_1 \mid \widetilde{\mathbf{u}}_t = \mathbf{u}_t] = \bar{\mathbf{q}}_t + (1-t)V^\star(\bar{\mathbf{q}}_t, \mathbf{y}_t, t). \tag{37}$$

Therefore, the trained velocity-prediction network $V_\theta(\bar{\mathbf{q}}_t, \mathbf{y}_t, t)$ provides an estimate of the conditional mean $\mathbb{E}[\widetilde{\mathbf{x}}_1 \mid \widetilde{\mathbf{u}}_t = \mathbf{u}_t]$, denoted by $\widehat{\mathbf{x}}_1(\mathbf{u}_t, t)$, as follows:

$$\widehat{\mathbf{x}}_1(\mathbf{u}_t, t) = \bar{\mathbf{q}}_t + (1-t)V_\theta(\bar{\mathbf{q}}_t, \mathbf{y}_t, t), \tag{38}$$

as stated earlier in Eq. (14). The joint-state sampling update can now be implemented with this estimate.

For initialization of this sampling procedure, we define the network input at $t = 0$ by the right-hand limit of $\bar{\mathbf{q}}_t$, i.e., $\lim_{t\to 0^+} \bar{\mathbf{q}}_t$. To evaluate this limit, we rewrite Eq. (35) as

$$\begin{aligned}\bar{\mathbf{q}}_t &= (\sigma^2 t^2 \mathbf{I} + \mathbf{A}^T\mathbf{A})^{-1}(\sigma^2 t^2 \mathbf{x}_t + t\mathbf{A}^T\mathbf{y}_t) \\ &= (\sigma^2 t^2 \mathbf{I} + \mathbf{A}^T\mathbf{A})^{-1}\big[(\sigma^2 t^2 \mathbf{I} + \mathbf{A}^T\mathbf{A})\mathbf{x}_t \\ &\quad - \mathbf{A}^T\mathbf{A}\mathbf{x}_t + t\mathbf{A}^T\mathbf{y}_t\big] \\ &= \mathbf{x}_t - (\sigma^2 t^2 \mathbf{I} + \mathbf{A}^T\mathbf{A})^{-1}\mathbf{A}^T(\mathbf{A}\mathbf{x}_t - t\mathbf{y}_t). \end{aligned} \tag{39}$$

Based on the SVD representation of the matrix $(\sigma^2 t^2 \mathbf{I} + \mathbf{A}^T\mathbf{A})^{-1}\mathbf{A}^T$ used for Tikhonov regularization with parameter $\sigma^2 t^2$, or equivalently on Tikhonov's regularization theorem [40, Th. 4.3], we have

$$\lim_{t\to 0^+} (\sigma^2 t^2 \mathbf{I} + \mathbf{A}^T\mathbf{A})^{-1}\mathbf{A}^T = \mathbf{A}^+,$$

where $\mathbf{A}^+$ denotes the Moore–Penrose pseudoinverse of $\mathbf{A}$. Using this to evaluate the limit in Eq. (39), we obtain the network input at $t = 0$ as

$$\lim_{t\to 0^+} \bar{\mathbf{q}}_t = (\mathbf{I} - \mathbf{A}^+\mathbf{A})\mathbf{x}_0. \tag{40}$$

Hence, the initial input of the network is the projection of the random sample $\mathbf{x}_0$ onto the nullspace of $\mathbf{A}$. When $\mathbf{A}$ is not full column rank, like in inpainting and super-resolution problems, this projection enables the initial network input to retain the values of $\mathbf{x}_0$ only at the locations of missing observations while setting the remaining pixel values to zero, therefore allowing the capacity of the network to focus more on the missing information. In addition, when $\mathbf{A}$ has full column rank, like in denoising problems, the initial network input becomes zero everywhere, providing the same initialization as with evaluating Eq. (35) at $t = 0$.

The resulting sampling procedure is summarized in Algorithm 2 and illustrated in Fig. 1. In the initialization step, the only hyperparameter $N$, representing the number of inference steps, is set, and the acquired noisy measurement $\mathbf{y}_0$ and an initial random sample $\mathbf{x}_0 \sim \mathcal{N}(\mathbf{0}, \mathbf{I})$ are assigned as $\mathbf{y}^{(0)} = \mathbf{y}_0$ and $\mathbf{x}^{(0)} = \mathbf{x}_0$, respectively. These are used to form the initial joint-state sample $\mathbf{u}^{(0)}$, and then $\bar{\mathbf{q}}^{(0)}$ via Eq. (40). During Dual Flow sampling, at iteration $k$, the network input $\bar{\mathbf{q}}^{(k)}$ is computed by evaluating Eq. (35) at $t = t_k$ with the available joint-state sample $\mathbf{u}^{(k)} = [(\mathbf{x}^{(k)})^T \ (\mathbf{y}^{(k)})^T]^T$. The estimated conditional mean using Eq. (38) at $t = t_k$ with $\bar{\mathbf{q}}^{(k)}$ and $\mathbf{y}^{(k)}$ is then utilized in the joint-state sampling update in Eq. (11).

**Algorithm 1** Training

**Input:** $\mathbf{A}, \sigma$

```
1: repeat
2:    t ~ U(0,1),  x_1 ~ p_data,  x_0 ~ N(0, I),  n ~ N(0, σ²I)
3:    u_1 ← [x_1; A x_1],   u_0 ← [x_0; A x_1 + n]
4:    u_t ← t u_1 + (1 − t) u_0
5:    q̄_t ← (t² I + σ⁻² AᵀA)⁻¹ [t² I   t σ⁻² Aᵀ] u_t
6:    Update θ based on Eq. (34)
7: until convergence
```

1: **repeat**
2: $t \sim \mathcal{U}(0,1),\ \mathbf{x}_1 \sim p_{\text{data}},\ \mathbf{x}_0 \sim \mathcal{N}(\mathbf{0},\mathbf{I}),\ \mathbf{n} \sim \mathcal{N}(\mathbf{0},\sigma^2\mathbf{I})$
3: $\mathbf{u}_1 \leftarrow \begin{bmatrix}\mathbf{x}_1 \\ \mathbf{A}\mathbf{x}_1\end{bmatrix},\quad \mathbf{u}_0 \leftarrow \begin{bmatrix}\mathbf{x}_0 \\ \mathbf{A}\mathbf{x}_1 + \mathbf{n}\end{bmatrix}$
4: $\mathbf{u}_t \leftarrow t\mathbf{u}_1 + (1-t)\mathbf{u}_0$
5: $\bar{\mathbf{q}}_t \leftarrow \left(t^2\mathbf{I} + \sigma^{-2}\mathbf{A}^T\mathbf{A}\right)^{-1}\left[t^2\mathbf{I} \quad t\sigma^{-2}\mathbf{A}^T\right]\mathbf{u}_t$
6: Update $\theta$ based on Eq. (34)
7: **until** convergence

**Algorithm 2** Dual Flow

**Input:** $\mathbf{y}_0, \mathbf{A}, \sigma, V_\theta$

**Hyperparameter:** $N$

1: $\Delta \leftarrow 1/N, \quad \mathbf{x}_0 \sim \mathcal{N}(\mathbf{0},\mathbf{I})$
2: $\mathbf{x}^{(0)} \leftarrow \mathbf{x}_0, \quad \mathbf{y}^{(0)} \leftarrow \mathbf{y}_0$
3: $\mathbf{u}^{(0)} \leftarrow \left[(\mathbf{x}^{(0)})^T \quad (\mathbf{y}^{(0)})^T\right]^T, \quad \bar{\mathbf{q}}^{(0)} \leftarrow (\mathbf{I} - \mathbf{A}^+\mathbf{A})\mathbf{x}^{(0)}$
4: **for** $k = 0$ to $N-1$ **do**
5: $t_k \leftarrow k\Delta$
6: **if** $k > 0$ **then**
7: $\bar{\mathbf{q}}^{(k)} \leftarrow \left(t_k^2\mathbf{I} + \sigma^{-2}\mathbf{A}^T\mathbf{A}\right)^{-1}\left[t_k^2\mathbf{I} \quad t_k\sigma^{-2}\mathbf{A}^T\right]\mathbf{u}^{(k)}$
8: $\widehat{\mathbf{x}}_1(\mathbf{u}^{(k)}, t_k) \leftarrow \bar{\mathbf{q}}^{(k)} + (1-t_k)V_\theta(\bar{\mathbf{q}}^{(k)}, \mathbf{y}^{(k)}, t_k)$
9: $\mathbf{u}^{(k+1)} \leftarrow \mathbf{u}^{(k)} + \frac{\Delta}{1-t_k}\left(\begin{bmatrix}\mathbf{I} \\ \mathbf{A}\end{bmatrix}\widehat{\mathbf{x}}_1(\mathbf{u}^{(k)}, t_k) - \mathbf{u}^{(k)}\right)$
10: **return** $\mathbf{x}^{(N)}$

## IV. Numerical Results

We evaluate the developed method on five linear inverse problems: denoising, Gaussian deblurring, super-resolution, random inpainting, and box inpainting. To this end, we first fine-tune unconditional velocity-prediction networks pretrained on human-face and cat images for each problem except denoising. We then conduct comprehensive experiments on both datasets, comparing the proposed method with recent approaches in terms of distortion metrics, perceptual quality, and computational efficiency. We also provide detailed analyses of the evolution of the image- and measurement-domain states during reverse sampling, the effects of the number of inference steps, robustness to varying degradation levels and model mismatch, reconstruction diversity, and uncertainty estimation.

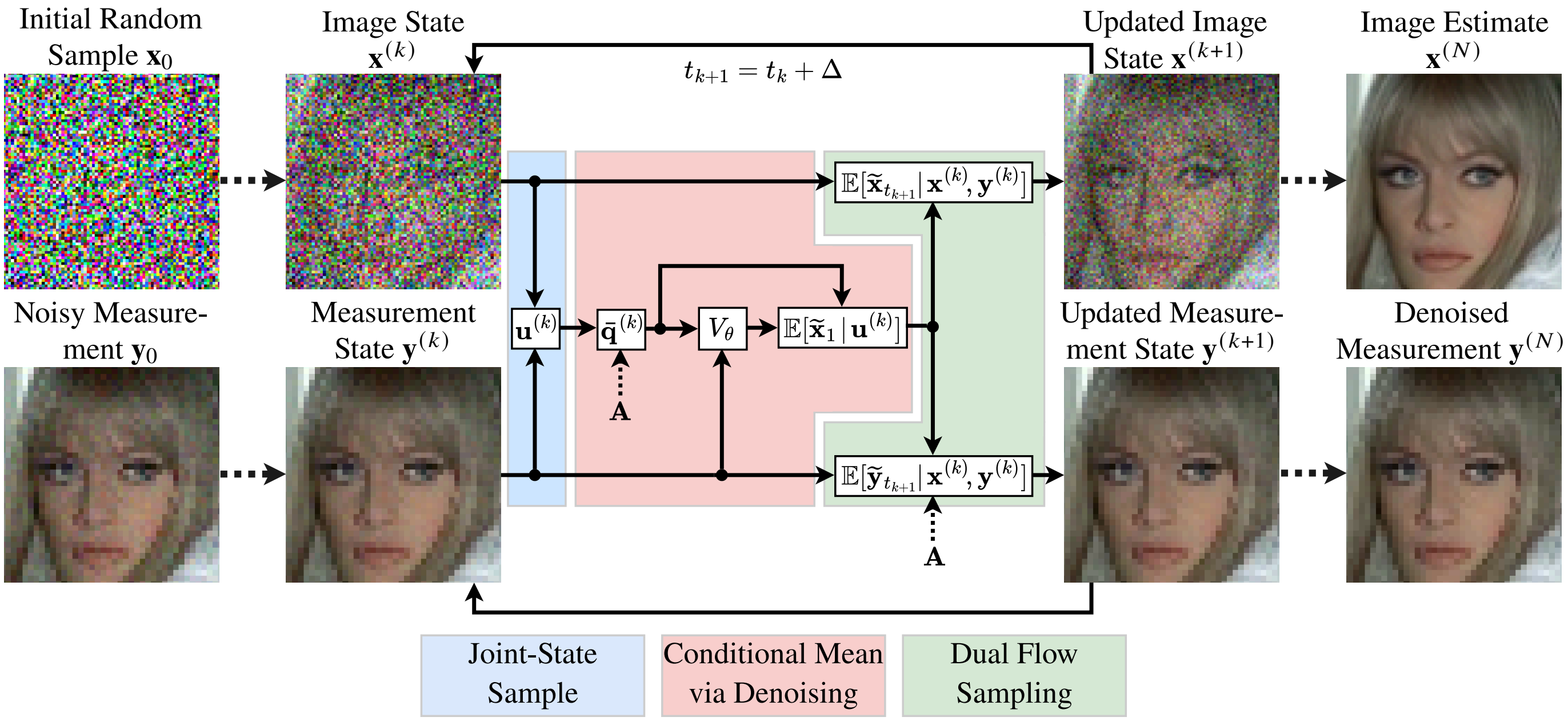

Fig. 1. Overview of Dual Flow. The reverse sampling is initialized with the acquired noisy measurement $\mathbf{y}_0$ and a random sample $\mathbf{x}_0 \sim \mathcal{N}(\mathbf{0}, \mathbf{I})$. At iteration $k$, the joint-state sample $\mathbf{u}^{(k)} = [(\mathbf{x}^{(k)})^T \; (\mathbf{y}^{(k)})^T]^T$ is formed from the image state sample $\mathbf{x}^{(k)}$ and measurement state sample $\mathbf{y}^{(k)}$. The joint-state sample and forward operator $\mathbf{A}$ are used to compute the scaled equivalent observation $\bar{\mathbf{q}}^{(k)}$, which, together with $\mathbf{y}^{(k)}$, is provided as input to the velocity-prediction network $V_\theta$. Subsequently, the network output and $\bar{\mathbf{q}}^{(k)}$ are used to estimate the conditional mean $\mathbb{E}[\widetilde{\mathbf{x}}_1 \,|\, \widetilde{\mathbf{u}}_{t_k} = \mathbf{u}^{(k)}]$. This estimate is then used in the Dual Flow sampling updates, which are based on the conditional expectations of the next image state, $\mathbb{E}[\widetilde{\mathbf{x}}_{t_{k+1}} \,|\, \widetilde{\mathbf{u}}_{t_k} = \mathbf{u}^{(k)}]$, and the next measurement state, $\mathbb{E}[\widetilde{\mathbf{y}}_{t_{k+1}} \,|\, \widetilde{\mathbf{u}}_{t_k} = \mathbf{u}^{(k)}]$. Finally, at $t_N = 1$, Dual Flow outputs the estimate of the unknown image $\mathbf{x}^{(N)}$ and its noise-free measurement $\mathbf{y}^{(N)} = \mathbf{A}\mathbf{x}^{(N)}$.

## A. Experimental Setup

We follow the experimental procedure of PnP-Flow [26]. We use CelebA [41] with human-face images resized to $128 \times 128$ and AFHQ-Cat [42] with cat images resized to $256 \times 256$. All images are normalized to $[-1, 1]$. We use the standard training, validation, and test splits of CelebA. Since AFHQ-Cat does not have a validation split, we reserve 32 images from its test set for validation. The acquired noisy measurements $\mathbf{y}_0$ are generated according to Eq. (1), with additive white Gaussian noise $\mathbf{n} \sim \mathcal{N}(\mathbf{0}, \sigma^2\mathbf{I})$.

We consider five linear inverse problems: denoising ($\sigma = 0.2$); Gaussian deblurring with a $61 \times 61$ kernel ($\sigma_b = 1$ for CelebA and $\sigma_b = 3$ for AFHQ-Cat, where $\sigma_b$ denotes the standard deviation of the blur kernel, and $\sigma = 0.05$); super-resolution with $2\times$ and $4\times$ decimation for CelebA and AFHQ-Cat, respectively ($\sigma = 0.05$); random inpainting with 70% missing pixels ($\sigma = 0.01$); and box inpainting with centered $40 \times 40$ and $80 \times 80$ missing regions for CelebA and AFHQ-Cat, respectively ($\sigma = 0.05$). We evaluate each inverse problem on 100 test images per dataset.

We compare the developed method with the recent flow-based algorithms PnP-Flow [26], Flower [27], OT-ODE [43], D-Flow [44], and Flow-Priors [45], as well as the plug-and-play method PnP-GS [46] and the diffusion-based method PnP-Diff [47]. For CelebA and AFHQ-Cat, all flow-based algorithms use the same pretrained, dataset-specific unconditional velocity-prediction networks, while PnP-GS uses separately trained denoisers. We obtain these networks and their pretrained weights directly from PnP-Flow [26]. Both the velocity-prediction networks and PnP-GS denoisers employ the same U-Net architecture [48], based on the implementations in [19], [49]. The velocity-prediction networks were trained using Mini-Batch Optimal-Transport Flow Matching [38] with a standard Gaussian latent distribution $\mathcal{N}(\mathbf{0}, \mathbf{I})$. Training was performed using the Adam optimizer with a learning rate of $10^{-4}$ for 200 and 400 epochs on CelebA and AFHQ-Cat, respectively, with corresponding batch sizes of 128 and 64. For PnP-Diff, following PnP-Flow [26], we use a pretrained diffusion model from [50], trained on the FFHQ dataset [51], since training a diffusion model with the same U-Net yielded poor results due to insufficient model capacity.

## B. Implementation Details

For denoising, we use the same pretrained unconditional velocity-prediction networks as the other flow-based methods, without additional training. In this case, since $\mathbf{A} = \mathbf{I}$, the noise term $\widetilde{\bar{\mathbf{n}}}_t$ in Eq. (31) is uncorrelated Gaussian with standard deviation $(1-t)/\sqrt{t^2 + \sigma^{-2}}$. The equivalent observation $\mathbf{q}_t = (t\mathbf{x}_t + \sigma^{-2}\mathbf{y}_t)/(t^2 + \sigma^{-2})$ can therefore be expressed as $\mathbf{x}_1 + \bar{\mathbf{n}}_t$, where $\bar{\mathbf{n}}_t$ denotes a given value of $\widetilde{\bar{\mathbf{n}}}_t$. To use the pretrained unconditional network, we scale the equivalent observation $\mathbf{q}_t$ to match the straight-line interpolation in Eq. (2). Specifically, scaling $\mathbf{q}_t$ by $\alpha_t$ gives $\alpha_t \mathbf{q}_t = \alpha_t \mathbf{x}_1 + \alpha_t \bar{\mathbf{n}}_t$. We then choose $\alpha_t$ such that the standard deviation of the scaled noise $\alpha_t \bar{\mathbf{n}}_t$ equals $1 - \alpha_t$, as required by the straight-line interpolation. This gives $\alpha_t (1-t)/\sqrt{t^2 + \sigma^{-2}} = 1 - \alpha_t$, yielding $\alpha_t = [1 + (1-t)/\sqrt{t^2 + \sigma^{-2}}]^{-1}$. The pretrained network is therefore evaluated at $(\alpha_t \mathbf{q}_t, \alpha_t)$ to estimate the conditional mean at the transformed flow time $\alpha_t$, for all $t \in [0, 1)$.

For the remaining inverse problems, we fine-tune the same pretrained unconditional velocity-prediction networks used by the other flow-based methods in our comparison. To incorporate the measurement-domain state $\mathbf{y}_t$ into the network, we expand the input convolutional layer from three to six channels. In doing so, we retain the pretrained weights for the original channels and initialize the additional weights to zero. The resulting network takes $(\bar{\mathbf{q}}_t, \mathbf{y}_t, t)$ as input. Here, $\mathbf{y}_t$ corresponds to the blurred measurement state for Gaussian deblurring and the masked measurement state for inpainting. For super-resolution, $\mathbf{y}_t$ is upsampled to the image dimensions using nearest-neighbor interpolation.

For all problems except denoising, these networks are fine-tuned using the same optimizer and batch-size settings as PnP-Flow [26]. For CelebA, fine-tuning is performed for approximately 50 epochs for all problems except random inpainting, for which approximately 150 epochs are used. For AFHQ-Cat, fine-tuning is performed for approximately 600 epochs for all problems. All experiments are conducted on a single NVIDIA A100 80 GB GPU.

We next describe how the network input $\bar{\mathbf{q}}_t$ in Eq. (35) is computed without explicit matrix inversion. For $t > 0$, $\mathbf{A}^T\mathbf{A}$ is diagonal for inpainting and super-resolution, allowing the network input to be computed by elementwise division in Eq. (35). For Gaussian deblurring, $\mathbf{A}^T\mathbf{A}$ is diagonalized by the two-dimensional discrete Fourier transform (DFT) under circular convolution. Therefore, we compute the network input by elementwise division in the Fourier domain, followed by the inverse DFT. For initialization at $t = 0$, Eq. (40) shows that the network input is the projection of $\mathbf{x}_0$ onto the nullspace of $\mathbf{A}$. For inpainting and super-resolution, this projection retains the values of $\mathbf{x}_0$ at unobserved pixel locations and sets the remaining pixels to zero. For Gaussian deblurring, we approximate the projection in the Fourier domain. Specifically, letting $\mathbf{F}$ denote the unitary DFT matrix, $\mathbf{FAF}^H$ is diagonal. We retain only the coefficients of $\mathbf{Fx}_0$ corresponding to diagonal entries of $\mathbf{FAF}^H$ whose magnitudes are at most $\sigma \times 10^{-6}$ and set the remaining coefficients to zero. The network input at $t = 0$ is then obtained by applying the inverse DFT.

### *C. Comparison With Other Methods*

For each competing method, we use the corresponding hyperparameter settings reported in the comparative evaluations of PnP-Flow [26] and Flower [27]. The hyperparameters of the methods compared in PnP-Flow were selected through a grid search on the validation set to maximize peak signal-to-noise ratio (PSNR). Among the competing methods, we report two variants of PnP-Flow: PnP-Flow1 uses one velocity-network evaluation per denoising step, whereas PnP-Flow5 averages the outputs of five velocity-network evaluations per denoising step. We also report two variants of Flower: Flower1 uses a single reconstruction, whereas Flower5 averages five reconstructions.

The Dual Flow sampling procedure has no tunable hyperparameters other than the number of inference steps $N$. Since each reverse sampling step requires a single network evaluation, we consider two variants of our method with the same total number of network evaluations, denoted by $N_e$, which is the dominant factor in runtime. These variants are denoted by Dual Flow-T and Dual Flow-W, where T and W stand for Tall and Wide, respectively. Specifically, Dual Flow-T generates a single reconstruction using $N_e$ steps ($\Delta = 1/N_e$), whereas Dual Flow-W averages $K$ independently generated reconstructions, each using $N_e/K$ steps ($\Delta = K/N_e$). Thus, both variants require a total of $N_e$ network evaluations. For CelebA, we use $N_e = 18$ and $K = 6$ for all inverse problems. For AFHQ-Cat, we use $(N_e, K) = (40, 20)$ for denoising, $(40, 10)$ for Gaussian deblurring, $(40, 40)$ for super-resolution, and $(20, 10)$ for both inpainting problems.

We evaluate reconstruction quality using the distortion metrics peak signal-to-noise ratio (PSNR) and structural similarity index (SSIM), and perceptual quality using learned perceptual image patch similarity (LPIPS). Higher PSNR and SSIM values and lower LPIPS values indicate better performance. We also report average reconstruction time per image. Throughout Section IV, the values reported below the reconstructed images in the qualitative figures correspond to PSNR, SSIM, and LPIPS, respectively.

Table I presents the average reconstruction performance of the compared methods over 100 test images on CelebA. As in PnP-Flow [26], box-inpainting results are omitted (indicated by a dash, –) for PnP-GS, which is not designed for generative tasks, and PnP-Diff, whose diffusion model was pretrained on a different dataset. Dual Flow achieves state-of-the-art reconstruction performance, outperforming competing methods in terms of both distortion metrics and perceptual quality. In particular, Dual Flow-W achieves the highest PSNR for all five inverse problems and the highest SSIM for four of them, with a value very close to the best result for Gaussian deblurring. Dual Flow-T achieves the lowest LPIPS across all five problems. In terms of computational efficiency, Dual Flow requires substantially less reconstruction time than all competing methods, except PnP-GS for denoising.

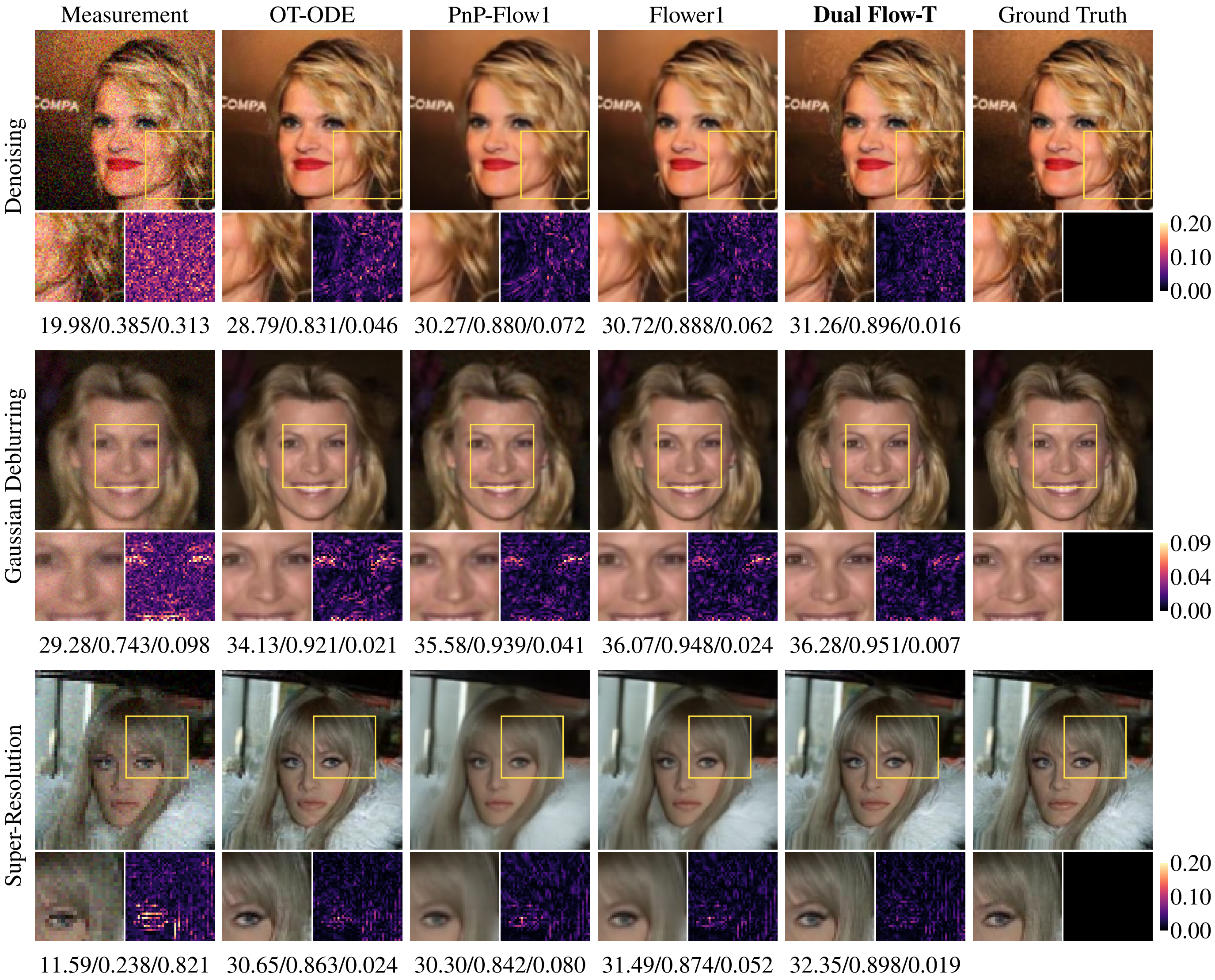

Fig. 2. Qualitative comparison of reconstruction results for denoising, Gaussian deblurring, and super-resolution on CelebA. Each result shows the full image, a magnified region, and a pixelwise absolute-error map of that region. The values below the images are PSNR, SSIM, and LPIPS, respectively.

The qualitative results in Figs. 2 and 3 show that Dual Flow-T better preserves fine textures and sharp details while producing fewer visible artifacts compared with the competing methods. In contrast to the other methods, our approach generally avoids the over-smoothing observed in their reconstructions. For a closer visual comparison, the figures also include magnified regions of the reconstructed images and the corresponding pixelwise absolute-error maps, computed by averaging the absolute differences between the ground-truth and reconstructed images over RGB channels. Under a common color scale for each inverse problem, the error maps of Dual Flow-T exhibit darker regions, indicating smaller reconstruction errors in the displayed examples. Overall, these results demonstrate that Dual Flow achieves substantial improvements in perceptual quality without sacrificing distortion performance, yielding a more favorable perception–distortion trade-off than competing methods.

We further evaluate the compared methods on AFHQ-Cat dataset over 100 test images. These additional results are provided in the supplementary material. Similar to CelebA, Dual Flow-W achieves the best or near-best PSNR and SSIM across all five inverse problems, while Dual Flow-T achieves the lowest LPIPS for all problems. Dual Flow also requires substantially less reconstruction time than all competing methods, except PnP-GS for denoising.

### D. *Evolution of the Joint States*

We next examine the evolution of the image- and measurement-domain state samples during Dual Flow sampling. Fig. 4 shows the intermediate image and measurement state samples generated by Dual Flow-T for super-resolution on CelebA.

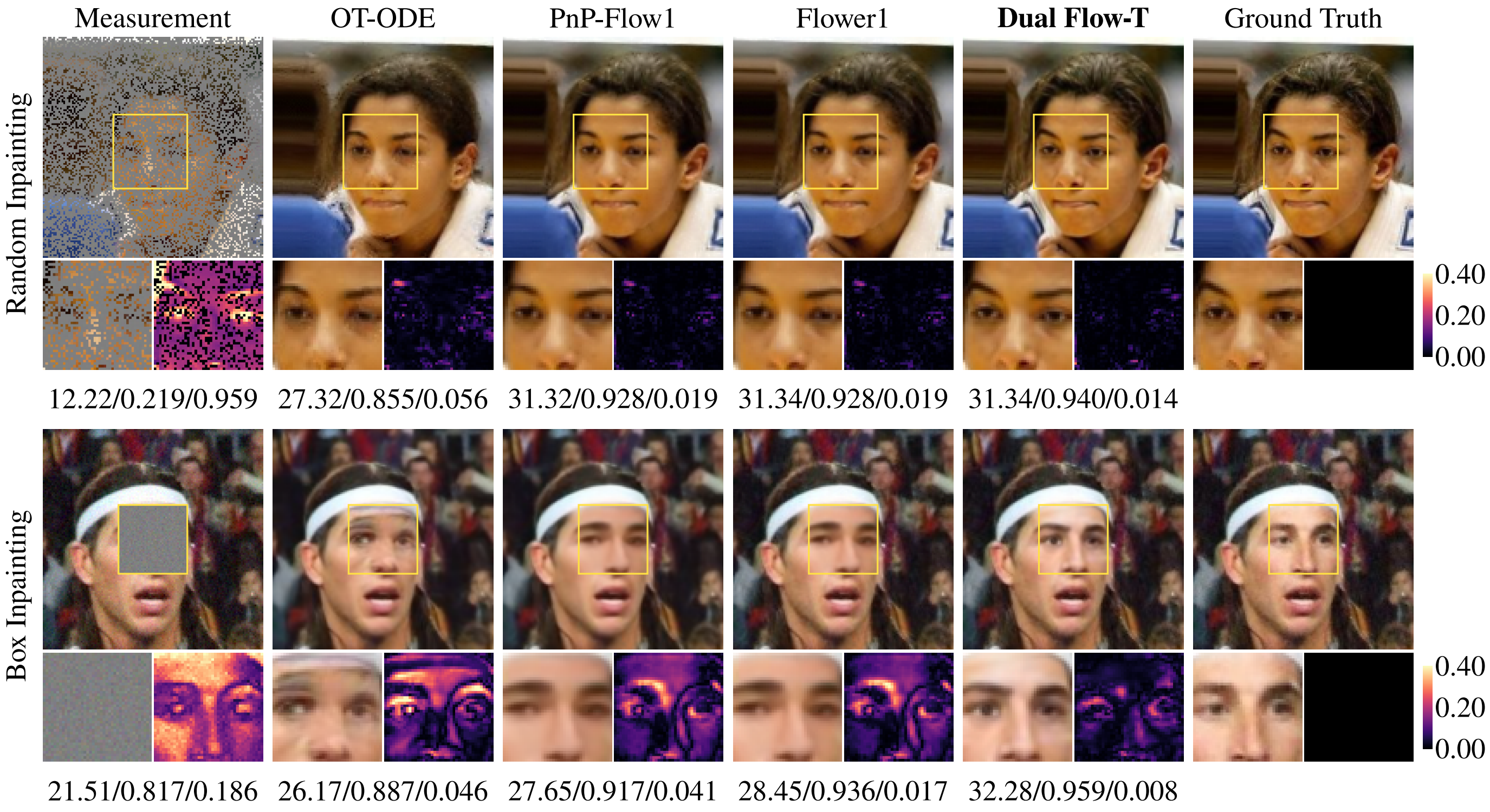

Fig. 3. Qualitative comparison of reconstruction results for random and box inpainting on CelebA. Each result shows the full image, a magnified region, and a pixelwise absolute-error map of that region. The values below the images are PSNR, SSIM, and LPIPS, respectively.

TABLE I
AVERAGE RECONSTRUCTION PERFORMANCE FOR DIFFERENT INVERSE PROBLEMS ON CELEBA OVER 100 TEST IMAGES. THE BEST AND SECOND-BEST RESULTS ARE SHOWN IN **BOLD** AND UNDERLINED, RESPECTIVELY.

| Method | Denoising | | | | Gaussian Deblurring | | | | Super-Resolution | | | | Random Inpainting | | | | Box Inpainting | | | |
|---|---|---|---|---|---|---|---|---|---|---|---|---|---|---|---|---|---|---|---|---|
| | PSNR ↑ | SSIM ↑ | LPIPS ↓ | Time (s) ↓ | PSNR ↑ | SSIM ↑ | LPIPS ↓ | Time (s) ↓ | PSNR ↑ | SSIM ↑ | LPIPS ↓ | Time (s) ↓ | PSNR ↑ | SSIM ↑ | LPIPS ↓ | Time (s) ↓ | PSNR ↑ | SSIM ↑ | LPIPS ↓ | Time (s) ↓ |
| Measurement | 20.00 | 0.348 | 0.372 | – | 27.83 | 0.740 | 0.127 | – | 10.26 | 0.183 | 0.827 | – | 11.98 | 0.198 | 1.028 | – | 22.26 | 0.742 | 0.214 | – |
| PnP-Flow1 | 31.82 | 0.905 | 0.045 | 0.99 | 34.49 | 0.936 | 0.040 | 0.99 | 31.07 | 0.902 | 0.045 | 1.01 | 33.11 | 0.945 | 0.018 | 0.99 | 30.44 | 0.933 | 0.037 | 0.98 |
| PnP-Flow5 | 32.31 | 0.911 | 0.056 | 4.88 | 34.80 | 0.940 | 0.047 | 4.87 | 31.50 | 0.907 | 0.055 | 4.90 | 34.07 | 0.954 | 0.021 | 4.83 | 31.06 | 0.940 | 0.043 | 4.83 |
| Flower1 | 32.28 | 0.913 | 0.034 | 1.03 | 34.98 | 0.947 | 0.026 | 1.77 | 32.34 | 0.922 | 0.034 | 1.08 | 33.14 | 0.945 | 0.017 | 1.05 | 31.13 | 0.945 | 0.023 | 1.03 |
| Flower5 | 33.15 | 0.926 | 0.038 | 5.06 | 35.67 | **0.954** | 0.032 | 8.87 | 33.09 | 0.931 | 0.039 | 5.40 | 34.03 | 0.954 | 0.019 | 5.25 | 31.83 | 0.952 | 0.023 | 5.15 |
| OT-ODE | 30.52 | 0.858 | 0.033 | 2.29 | 33.00 | 0.921 | 0.029 | 1.96 | 31.28 | 0.902 | 0.026 | 3.60 | 28.86 | 0.873 | 0.049 | 2.96 | 29.52 | 0.920 | 0.037 | 2.95 |
| D-Flow | 26.52 | 0.690 | 0.079 | 21.83 | 30.64 | 0.875 | 0.054 | 48.52 | 30.79 | 0.865 | 0.028 | 68.32 | 33.08 | 0.933 | 0.021 | 135.47 | 29.88 | 0.882 | 0.035 | 61.44 |
| Flow-Priors | 29.29 | 0.764 | 0.136 | 20.16 | 31.52 | 0.856 | 0.056 | 20.25 | 28.34 | 0.714 | 0.101 | 20.23 | 33.07 | 0.945 | 0.018 | 21.28 | 30.05 | 0.858 | 0.048 | 20.20 |
| PnP-GS | 32.26 | 0.902 | 0.035 | **0.05** | 32.85 | 0.885 | 0.046 | 0.95 | 30.44 | 0.881 | 0.082 | 0.57 | 27.89 | 0.820 | 0.106 | 0.66 | – | – | – | – |
| PnP-Diff | 31.19 | 0.885 | 0.061 | 0.80 | 32.77 | 0.912 | 0.061 | 0.80 | 31.49 | 0.894 | 0.033 | 3.71 | 31.86 | 0.919 | 0.024 | 0.82 | – | – | – | – |
| Dual Flow-T | 32.99 | 0.920 | **0.016** | 0.18 | 35.14 | 0.948 | **0.010** | **0.18** | 33.07 | 0.932 | **0.014** | **0.18** | 34.14 | 0.954 | **0.011** | **0.18** | 32.74 | 0.955 | **0.014** | **0.18** |
| Dual Flow-W | **33.54** | **0.931** | 0.020 | 0.18 | **35.73** | 0.953 | 0.016 | **0.18** | **34.28** | **0.946** | 0.021 | **0.18** | **35.73** | **0.965** | 0.014 | **0.18** | **33.71** | **0.961** | 0.015 | **0.18** |

Starting from the random sample $\mathbf{x}_0 \sim \mathcal{N}(\mathbf{0}, \mathbf{I})$ and the acquired noisy measurement $\mathbf{y}_0$, the image and measurement samples evolve jointly during sampling. As $t_k$ increases, the image state sample $\mathbf{x}^{(k)}$ progressively approaches the ground-truth image $\mathbf{x}_1$, with increasing PSNR and SSIM and decreasing LPIPS. Meanwhile, $\mathbf{y}^{(k)}$ evolves from the acquired noisy measurement $\mathbf{y}_0$ toward the corresponding noise-free measurement $\mathbf{y}_1 = \mathbf{A}\mathbf{x}_1$, as indicated by the decreasing relative measurement error $\varepsilon_y(t_k) = \|\mathbf{y}^{(k)} - \mathbf{y}_1\|_2 / \|\mathbf{y}_1\|_2$.

### E. Effect of the Number of Inference Steps

We further investigate the effect of the number of inference steps $N$ on reconstruction performance. To this end, we consider random inpainting on CelebA and evaluate the two variants of our method: Dual Flow-T, which generates a single reconstruction, and Dual Flow-W, which averages six independently generated reconstructions. For both variants, we use $N \in \{1, 3, 18, 30\}$ inference steps per reconstruction. The corresponding quantitative results, along with the total number of network evaluations $N_e$, are presented in Table II. Additional sample reconstruction results are provided in the supplementary document for visual evaluation.

As seen in Table II, even with $N = 1$, both variants outperform all competing methods in Table I in terms of PSNR and LPIPS, while Dual Flow-W also achieves the highest SSIM. Moreover, these quantitative results and the visual comparisons provided in the supplementary material suggest that increasing $N$ improves the perceptual quality of both Dual Flow-T and Dual Flow-W, as reflected by the generally decreasing LPIPS scores. Conversely, as $N$ decreases to 3, the reconstructions become smoother, with higher LPIPS values but improved distortion metrics (PSNR and SSIM). For fixed $N$ per reconstruction, Dual Flow-W achieves higher PSNR and SSIM by averaging multiple independently generated reconstructions, while generally yielding higher LPIPS than Dual Flow-T. These observations explain the complementary roles of the two Dual Flow variants. Under the same total number of network evaluations $N_e$, Dual Flow-W favors distortion metrics through reconstruction averaging with fewer steps per trajectory, whereas Dual Flow-T favors perceptual quality by allocating all $N_e$ network evaluations to a single trajectory.

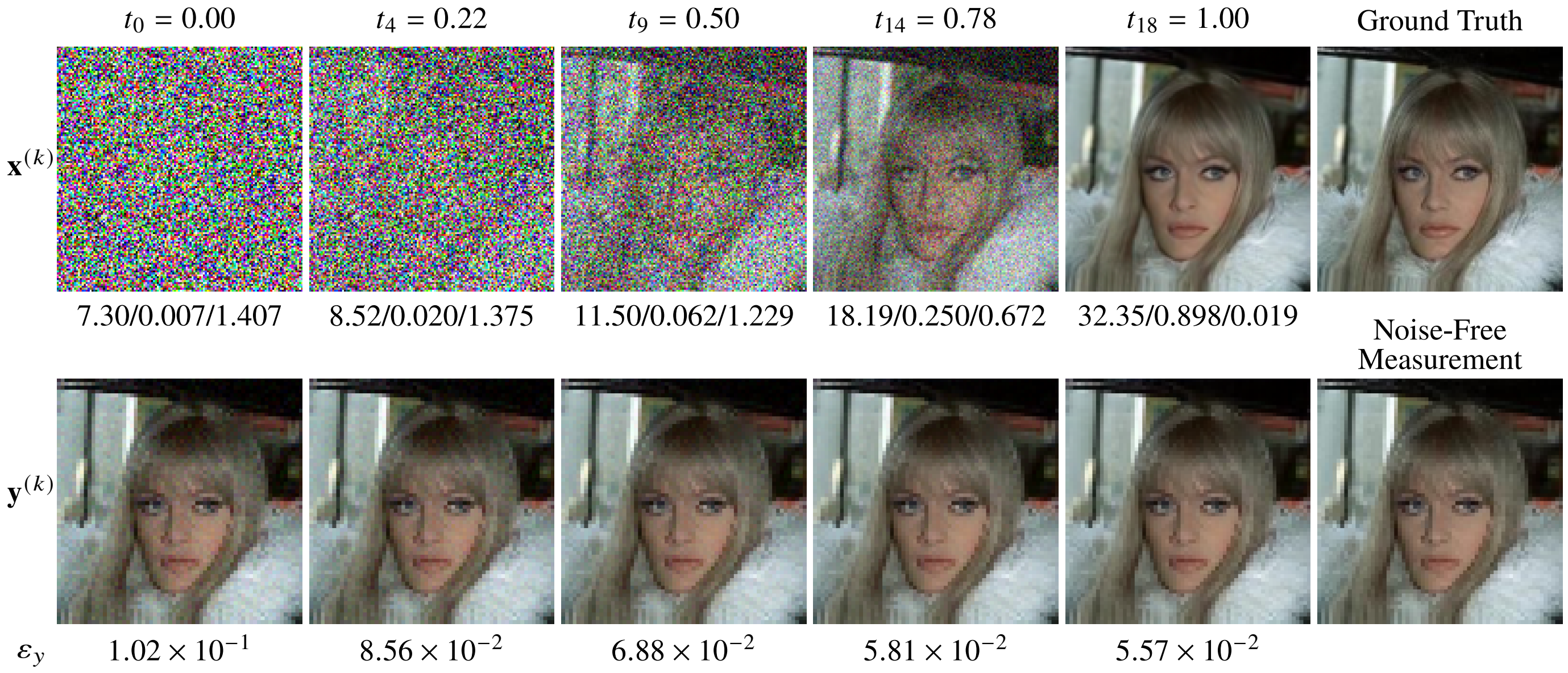

Fig. 4. Evolution of the image- and measurement-domain state samples during reverse sampling using Dual Flow-T. The first and second rows show the image state sample $\mathbf{x}^{(k)}$ and measurement state sample $\mathbf{y}^{(k)}$, respectively, at different time instants $t_k$. The last column shows the ground-truth image $\mathbf{x}_1$ and the corresponding noise-free measurement $\mathbf{y}_1 = \mathbf{A}\mathbf{x}_1$. The values below the first row are PSNR, SSIM, and LPIPS, respectively, while those below the second row are the relative measurement errors $\varepsilon_y(t_k) = \|\mathbf{y}^{(k)} - \mathbf{y}_1\|_2 / \|\mathbf{y}_1\|_2$.

TABLE II

Effect of the number of inference steps $N \in \{1, 3, 18, 30\}$ on random inpainting performance over 100 CelebA test images. The best results are shown in bold.

| Method | $N$ | $N_e$ | PSNR ↑ | SSIM ↑ | LPIPS ↓ |
|---|---|---|---|---|---|
| Dual Flow-T | 1 | 1 | 35.01 | 0.953 | 0.016 |
| | 3 | 3 | 35.20 | 0.961 | 0.013 |
| | 18 | 18 | 34.14 | 0.954 | **0.011** |
| | 30 | 30 | 33.92 | 0.952 | **0.011** |
| Dual Flow-W | 1 | 6 | 35.39 | 0.960 | 0.016 |
| | 3 | 18 | **35.73** | **0.965** | 0.014 |
| | 18 | 108 | 35.58 | **0.965** | 0.012 |
| | 30 | 180 | 35.52 | **0.965** | 0.012 |

### *F. Robustness to Varying Degradation Levels and Model Mismatch*

We now investigate the robustness of Dual Flow-T to variations in degradation settings on CelebA. Specifically, we consider different measurement noise levels for denoising and Gaussian blur kernel standard deviations for deblurring, without retraining the corresponding networks.

For denoising, we directly use the pretrained unconditional PnP-Flow network without fine-tuning, as described in Section IV-B. The derived formulation handles different measurement noise levels by evaluating the network at the transformed flow time $\alpha_t$. We test $\sigma \in \{0.1, 0.2, 0.4\}$, where $\sigma$ denotes the measurement noise standard deviation. As shown in Table III, although reconstruction performance degrades as the noise level increases, Dual Flow-T maintains high PSNR and SSIM and low LPIPS even at $\sigma = 0.4$. Consistent with these results, the visual results in the supplementary material illustrate that the reconstructions preserve the main facial structures and fine details even at this noise level.

For Gaussian deblurring, we investigate robustness to forward-operator mismatch between training and testing. The network is fine-tuned using a Gaussian blur kernel with standard deviation $\sigma_b = 1$. We then evaluate the same network at $\sigma_b \in \{0.8, 1.0, 1.2\}$, using the corresponding forward operator for each test setting. As shown in Table IV, both distortion and perceptual metrics degrade when the blur level differs from the training setting. Additional visual results provided in the supplementary document further illustrate that the reconstruction remains visually reasonable but appears over-sharpened at $\sigma_b = 0.8$, while more noticeable artifacts are observed at $\sigma_b = 1.2$.

TABLE III
AVERAGE DENOISING PERFORMANCE OF DUAL FLOW-T OVER 100 CELEBA TEST IMAGES FOR MEASUREMENT NOISE WITH STANDARD DEVIATION $\sigma \in \{0.1, 0.2, 0.4\}$.

| $\sigma$ | PSNR ↑ | SSIM ↑ | LPIPS ↓ |
|---|---|---|---|
| 0.10 | 36.48 | 0.957 | 0.007 |
| 0.20 | 32.99 | 0.920 | 0.016 |
| 0.40 | 29.73 | 0.866 | 0.032 |

TABLE IV
AVERAGE GAUSSIAN DEBLURRING PERFORMANCE OF DUAL FLOW-T ON 100 CELEBA TEST IMAGES FOR BLUR-KERNEL STANDARD DEVIATIONS $\sigma_b \in \{0.8, 1.0, 1.2\}$. THE NETWORK IS FINE-TUNED AT $\sigma_b = 1.0$.

| $\sigma_b$ | PSNR ↑ | SSIM ↑ | LPIPS ↓ |
|---|---|---|---|
| 0.80 | 30.67 | 0.877 | 0.038 |
| 1.00 | 35.14 | 0.948 | 0.010 |
| 1.20 | 23.82 | 0.511 | 0.389 |

Overall, these results indicate that the proposed method remains robust to challenging measurement noise levels in denoising. Here, the derived formulation in Section IV-B allows the pretrained unconditional network to handle varying noise levels without fine-tuning. However, as observed in Gaussian deblurring, the method remains sensitive to forward-operator mismatch when the forward operator used during testing differs from that used during training.

### G. Reconstruction Diversity and Uncertainty Estimation

Lastly, we assess Dual Flow-W's ability to generate diverse plausible reconstructions and characterize uncertainty, which cannot be captured by a single reconstruction. For a given noisy measurement, Dual Flow-W generates $K$ reconstructions from independent initial random samples, and we estimate pixelwise uncertainty from their variability.

Fig. 5 presents the uncertainty map computed from $K = 6$ samples generated by Dual Flow-W for the box-inpainting problem, along with three of the six reconstructions used to compute this map. The sample reconstructions exhibit variations in the missing region, particularly in fine facial details such as the eyebrow, where the estimated uncertainty is also higher. Thus, Dual Flow-W naturally provides diverse plausible reconstructions and pixelwise uncertainty estimates as byproducts of its sampling procedure. Additional details about this analysis are provided in the supplementary material.

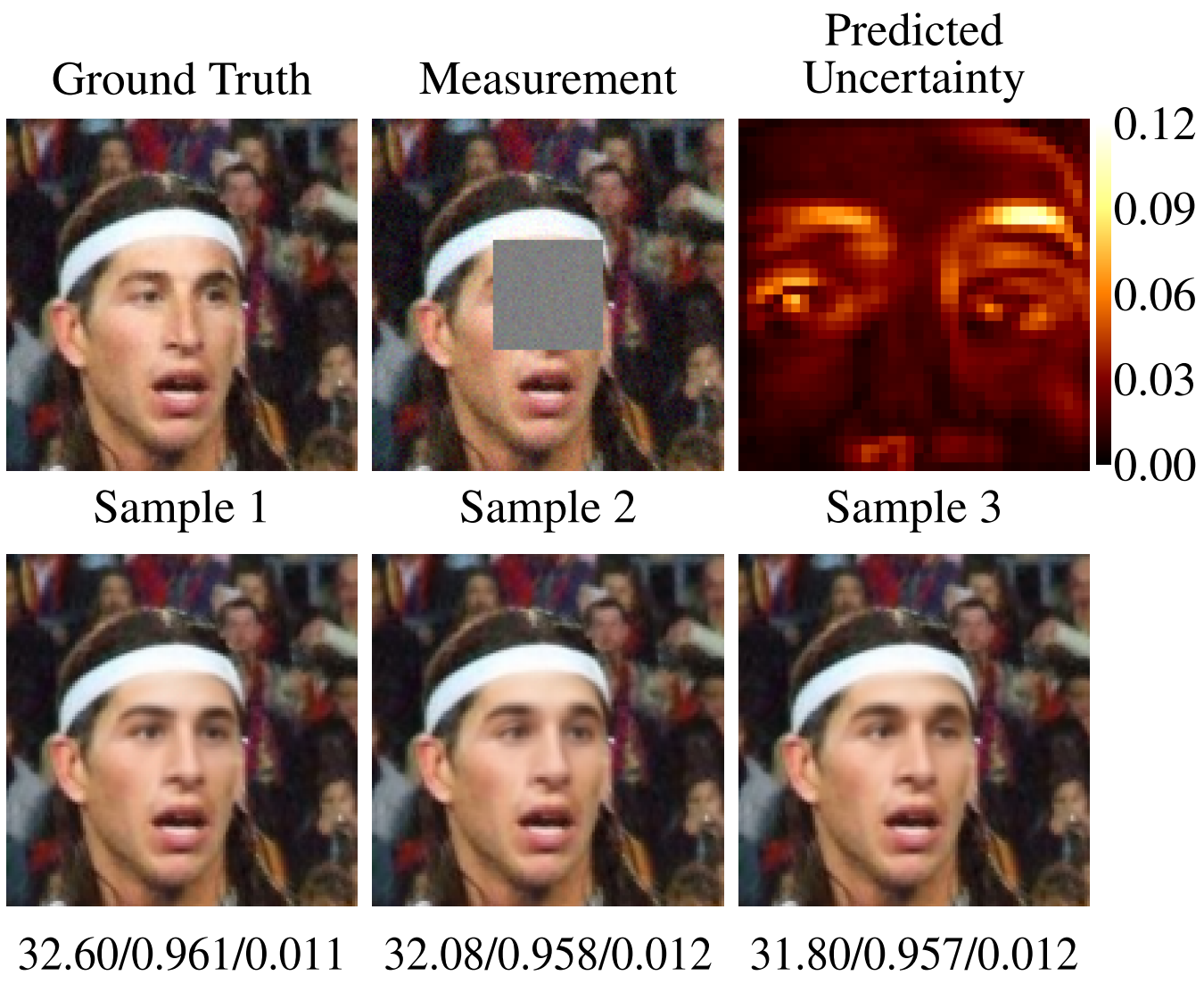

Fig. 5. Reconstruction diversity and uncertainty estimation using Dual Flow-W. The second row shows three representative reconstruction samples, while the predicted pixelwise uncertainty map in the missing box region is shown in the first row.

## V. CONCLUSION

In this paper, we developed Dual Flow, a novel joint flow model for solving linear inverse problems by jointly estimating the clean image and its corresponding noise-free measurement. The key idea is to jointly evolve the image and measurement states, rather than repeatedly using the acquired noisy measurement to condition or guide the reverse sampling process. This formulation enables image reconstruction and measurement denoising to be addressed as two coupled estimation tasks within a single sampling procedure. Our method does not require approximate measurement-conditioning strategies or any additional hyperparameters beyond the number of inference steps. By construction, the final image and measurement estimates are guaranteed to satisfy the forward model.

Extensive experiments across five linear inverse problems on two image datasets demonstrate that Dual Flow achieves state-of-the-art performance in both distortion metrics and perceptual quality, while requiring substantially less reconstruction time than competing methods. Visually, Dual Flow better preserves fine textures and sharp details while reducing over-smoothing and reconstruction artifacts. These findings demonstrate a more favorable perception–distortion trade-off than competing methods. Moreover, we showed that Dual Flow can generate diverse plausible reconstructions, enabling uncertainty estimation. Despite its strong reconstruction performance, Dual Flow remains sensitive to forward-operator mismatch between training and testing. Improving robustness to such mismatches is therefore an important future direction.

## Acknowledgments

This work was supported by the BAGEP Award of the Science Academy.

# *Dual Flow*: Joint Image–Measurement Interpolation for Inverse Problems

# Supplementary Material

Sevket Ege Yesilyurt and Figen S. Oktem

## Effect of the Number of Inference Steps and Comparison Between Dual Flow-T and Dual Flow-W

To analyze the effect of the number of inference steps, additional visual results are presented in Fig. 6.

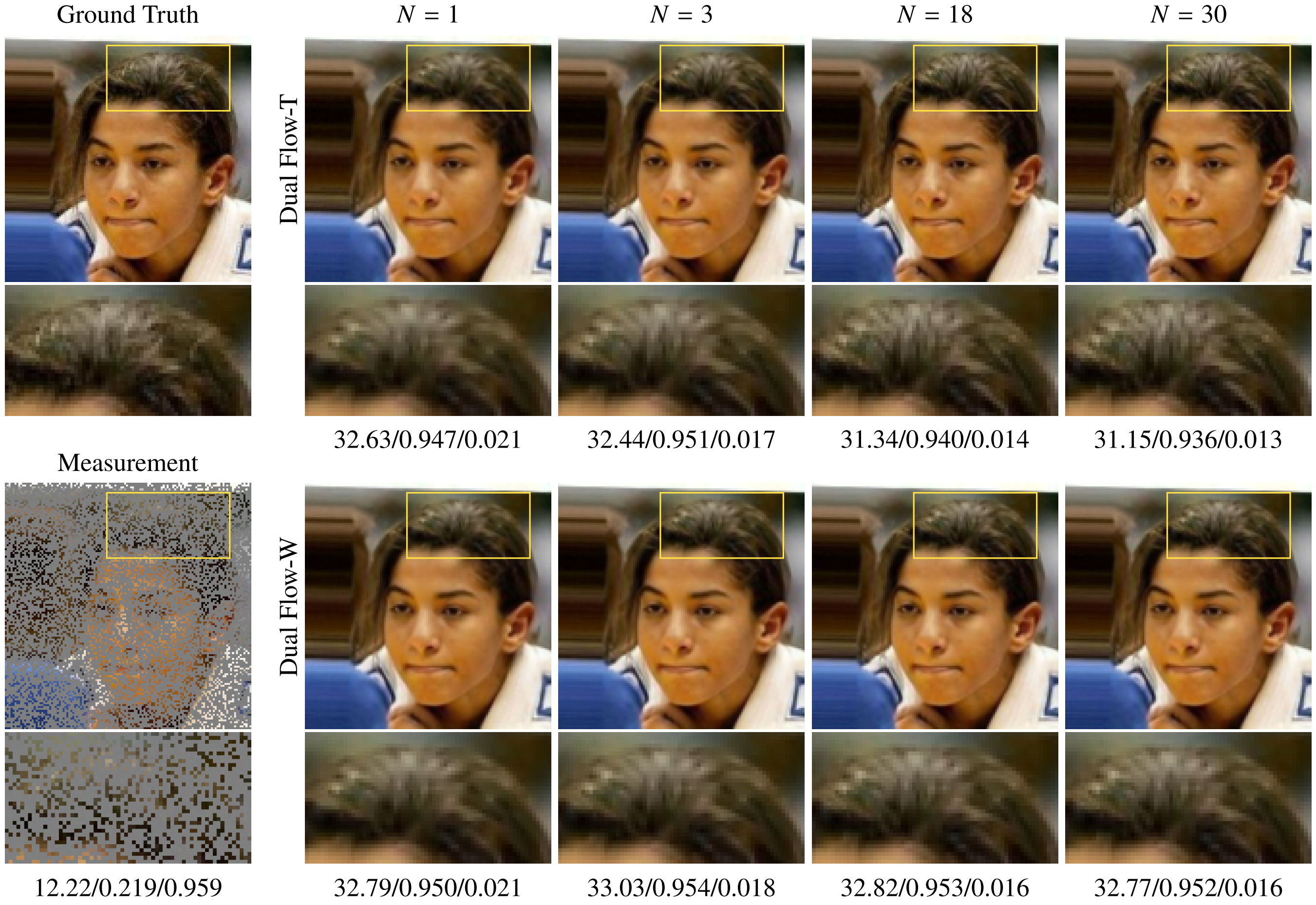

Fig. 6. Effect of the number of inference steps $N \in \{1, 3, 18, 30\}$ on random inpainting reconstructions on CelebA. The leftmost column shows the ground-truth image and the measurement. The first row shows Dual Flow-T, which generates a single reconstruction, whereas the second row shows Dual Flow-W, which averages six independently generated reconstructions, each using $N$ inference steps.

The visual comparisons in Fig. 6 demonstrate that, for both Dual Flow-T and Dual Flow-W, increasing the number of inference steps $N$ produces sharper reconstructions with finer details, as also reflected by decreasing LPIPS scores. Conversely, as $N$ decreases to 3, the reconstructions become smoother, with improved distortion metrics (PSNR and SSIM) but higher LPIPS values. For a fixed $N$ per reconstruction, Dual Flow-W produces smoother images with higher PSNR and SSIM by averaging six independently generated reconstructions, whereas Dual Flow-T produces a single sharper reconstruction with lower LPIPS. These observations further illustrate the complementary roles of the two variants: under the same total number of network evaluations $N_e$, Dual Flow-W favors distortion metrics through reconstruction averaging with fewer inference steps per reconstruction, while Dual Flow-T favors perceptual quality by producing a single reconstruction with $N = N_e$ inference steps.

### Robustness to Varying Degradation Levels and Model Mismatch

To visually assess the robustness of Dual Flow to varying degradation levels and forward-operator mismatch, sample reconstructions on CelebA are presented in Fig. 7.

For denoising, we directly use the pretrained unconditional network without additional fine-tuning, evaluating it at the transformed flow time as derived in Section IV-B of the main paper. This formulation allows Dual Flow to handle different measurement noise levels without requiring separate training for each noise level. As shown in Fig. 7(a), the reconstructions preserve facial structures and fine details even under a relatively high noise standard deviation of $\sigma = 0.4$. These results suggest the potential of Dual Flow to maintain reconstruction quality under more challenging degradation conditions in the absence of forward-operator mismatch between training and testing.

For Gaussian deblurring, we evaluate the network fine-tuned at $\sigma_b = 1.0$ under different blur-kernel standard deviations $\sigma_b \in \{0.8, 1.0, 1.2\}$, using the corresponding forward operator for each test setting without further fine-tuning. As shown in Fig. 7(b), deviations from the training blur level lead to degradation in both distortion metrics and perceptual quality. In particular, at $\sigma_b = 0.8$, the reconstruction remains visually reasonable but appears sharper than the ground-truth image, whereas at $\sigma_b = 1.2$, more noticeable structured artifacts are observed. These observations illustrate the sensitivity of Dual Flow to forward-operator mismatch between training and testing.

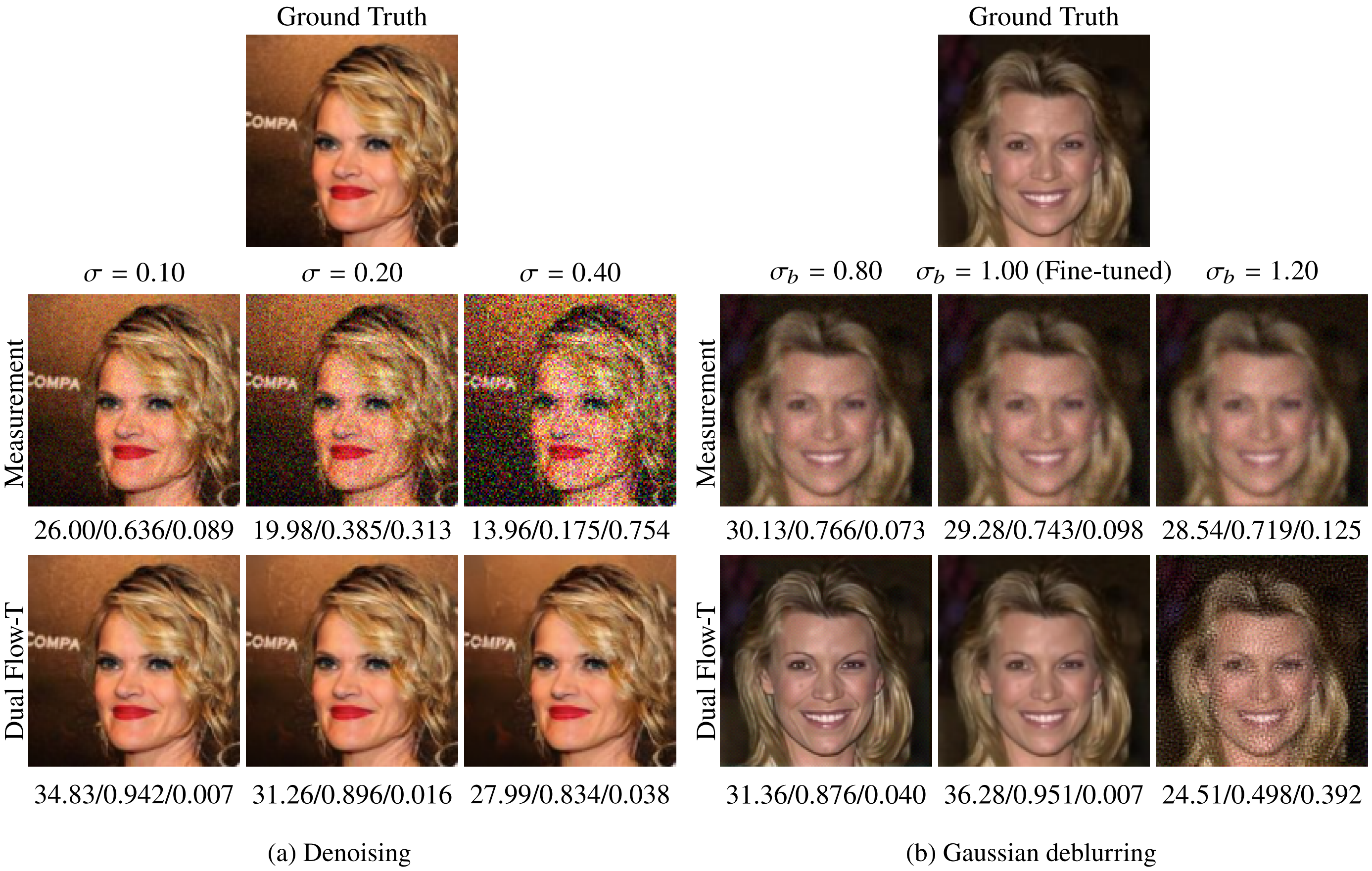

Fig. 7. Robustness results on CelebA using Dual Flow-T. (a) Denoising under measurement noise with standard deviation $\sigma \in \{0.1, 0.2, 0.4\}$, without fine-tuning the pretrained network. (b) Gaussian deblurring under forward-operator variations, where $\sigma_b$ denotes the standard deviation of the Gaussian blur kernel. The network is fine-tuned at $\sigma_b = 1.0$ and evaluated at $\sigma_b \in \{0.8, 1.0, 1.2\}$ without further fine-tuning.

## Reconstruction Diversity and Uncertainty Estimation

Beyond reconstruction accuracy, we investigate the reconstruction diversity and uncertainty estimation capabilities of Dual Flow-W. Point estimators, such as MAP or MMSE estimators, provide a single reconstruction. Although such estimates can be optimal under their respective estimation criteria, a single reconstruction cannot capture the full range of plausible solutions, thereby discarding information about uncertainty and reconstruction diversity.

Dual Flow-W addresses this limitation by generating $K$ reconstructions for the same acquired noisy measurement through joint-state sampling, starting from $K$ independent initial samples $\mathbf{x}_j^{(0)} \sim \mathcal{N}(\mathbf{0}, \mathbf{I})$, $j = 1, \ldots, K$. These reconstructions represent different plausible solutions supported by the learned joint flow model rather than random perturbations of a single estimate. The samples tend to agree in regions strongly constrained by the measurements, whereas regions with limited information due to the forward operator, noise, or missing observations may exhibit greater variability. This enables pixelwise uncertainty estimation directly from the generated reconstructions.

Given the $K$ final reconstructions $\{\mathbf{x}_j^{(N)}\}_{j=1}^{K}$, we estimate the pixelwise uncertainty by averaging the sample variances across RGB channels and taking the square root:

$$\hat{\sigma}[n] = \sqrt{\frac{1}{3(K-1)} \sum_{c=1}^{3} \sum_{j=1}^{K} \left( x_{j,c}^{(N)}[n] - \bar{x}_c^{(N)}[n] \right)^2}, \tag{41}$$

where $\bar{\mathbf{x}}^{(N)} = \frac{1}{K} \sum_{j=1}^{K} \mathbf{x}_j^{(N)}$ is the mean estimate, $c$ indexes the RGB channels, and $n$ denotes the pixel location.

This capability is relevant to box inpainting, where a contiguous image region is missing and multiple distinct reconstructions may be consistent with the same acquired noisy measurement. Fig. 5 of the main paper demonstrates three representative reconstructions and the predicted uncertainty map computed from the $K = 6$ samples generated by Dual Flow-W for a representative box-inpainting example from CelebA. The sample reconstructions exhibit variations in the missing region. These variations are particularly noticeable in fine facial details, such as the eyebrow, where the uncertainty map also exhibits higher standard deviations. These results illustrate that Dual Flow-W provides diverse plausible reconstructions and uncertainty estimates as an intrinsic byproduct of its sampling procedure, without requiring additional network evaluations beyond those used to generate the reconstruction samples.

## Numerical Results on AFHQ-Cat

We evaluate the compared methods on AFHQ-Cat dataset over 100 test images. Table V presents the corresponding quantitative results. Dual Flow-W achieves the highest PSNR for all five inverse problems except random inpainting, while obtaining the best or near-best SSIM across all five problems. Dual Flow-T achieves the lowest LPIPS across all five problems. Similar to the results on CelebA, Dual Flow requires substantially less reconstruction time than all competing methods, except PnP-GS for denoising.

The qualitative results in Figs. 8 and 9 show that Dual Flow-T preserves fine textures and sharp details while introducing fewer artifacts and avoiding the oversmoothing observed in competing methods. These results further demonstrate that Dual Flow-T provides a better balance between perceptual quality and distortion performance compared with competing methods.

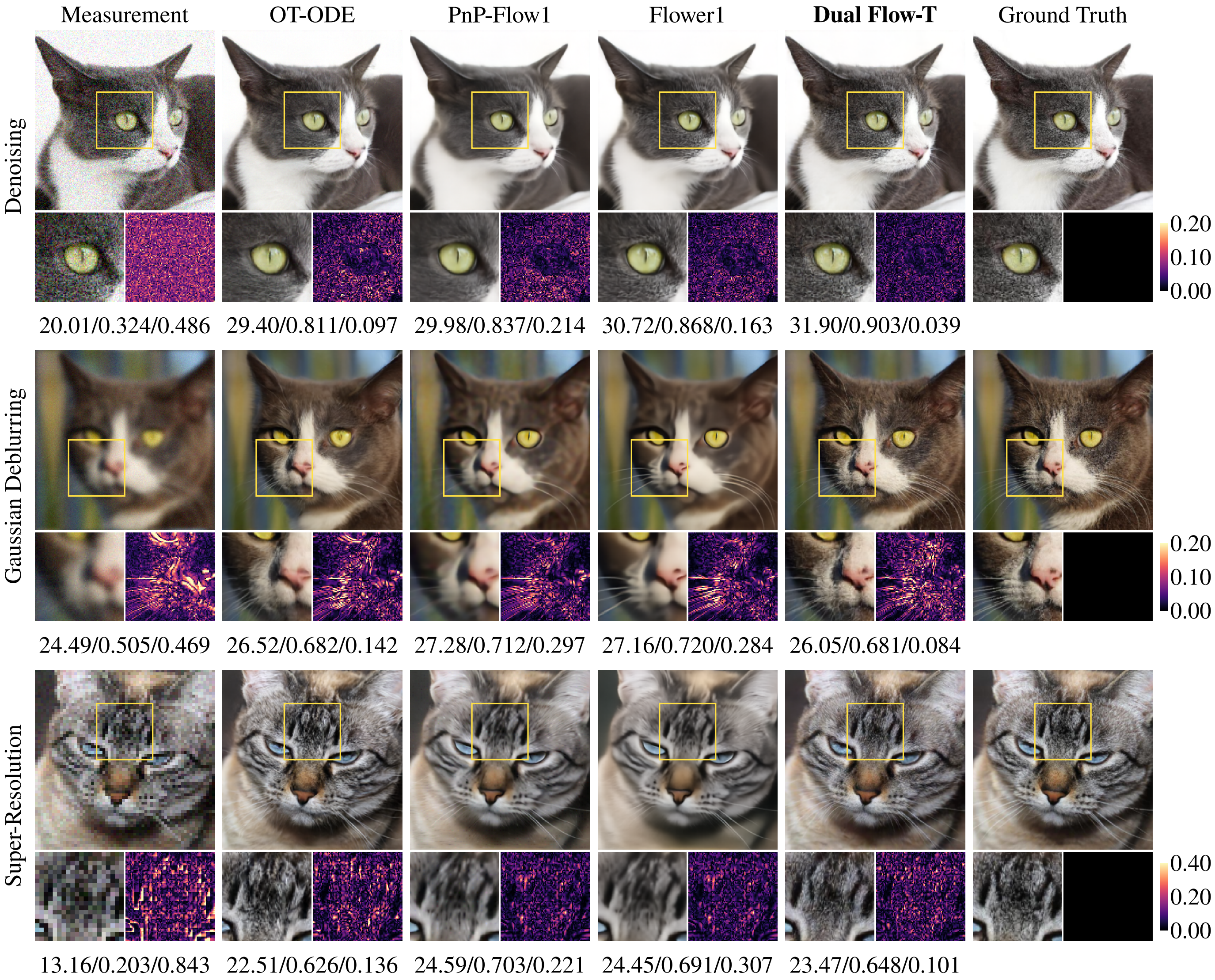

Fig. 8. Qualitative comparison of reconstruction results for denoising, Gaussian deblurring, and super-resolution on AFHQ-Cat. Each result shows the full image, a magnified region, and a pixelwise absolute-error map of that region. The values below the images are PSNR, SSIM, and LPIPS, respectively.

TABLE V

AVERAGE RECONSTRUCTION PERFORMANCE FOR DIFFERENT INVERSE PROBLEMS ON 100 AFHQ-CAT TEST IMAGES. THE BEST AND SECOND-BEST RESULTS ARE SHOWN IN **BOLD** AND UNDERLINED, RESPECTIVELY.

| Method | Denoising | | | | Gaussian Deblurring | | | | Super-Resolution | | | | Random Inpainting | | | | Box Inpainting | | | |
|---|---|---|---|---|---|---|---|---|---|---|---|---|---|---|---|---|---|---|---|---|
| | PSNR ↑ | SSIM ↑ | LPIPS ↓ | Time (s) ↓ | PSNR ↑ | SSIM ↑ | LPIPS ↓ | Time (s) ↓ | PSNR ↑ | SSIM ↑ | LPIPS ↓ | Time (s) ↓ | PSNR ↑ | SSIM ↑ | LPIPS ↓ | Time (s) ↓ | PSNR ↑ | SSIM ↑ | LPIPS ↓ | Time (s) ↓ |
| Measurement | 20.00 | 0.296 | 0.529 | – | 24.38 | 0.528 | 0.437 | – | 11.77 | 0.218 | 0.872 | – | 13.44 | 0.221 | 1.074 | – | 21.72 | 0.730 | 0.214 | – |
| PnP-Flow1 | 31.59 | 0.867 | 0.140 | 0.98 | 28.65 | 0.777 | 0.289 | 4.90 | 27.49 | 0.778 | 0.159 | 5.30 | 33.72 | 0.924 | 0.038 | 2.12 | 26.69 | 0.897 | 0.110 | 0.98 |
| PnP-Flow5 | 31.84 | 0.868 | 0.171 | 4.87 | 28.96 | 0.783 | 0.318 | 24.36 | 27.96 | **0.790** | 0.169 | 24.77 | **34.48** | **0.933** | 0.045 | 9.92 | 27.20 | 0.901 | 0.127 | 4.88 |
| Flower1 | 32.09 | 0.882 | 0.114 | 1.09 | 28.63 | 0.780 | 0.267 | 9.17 | 26.68 | 0.757 | 0.266 | 6.18 | 33.69 | 0.923 | 0.042 | 2.40 | 26.95 | 0.915 | 0.063 | 1.10 |
| Flower5 | 32.73 | 0.893 | 0.127 | 5.41 | 29.04 | 0.788 | 0.281 | 46.07 | 27.06 | 0.766 | 0.277 | 30.84 | 34.41 | 0.932 | 0.048 | 11.93 | 27.42 | 0.923 | 0.067 | 5.47 |
| OT-ODE | 30.44 | 0.818 | 0.081 | 2.19 | 27.76 | 0.732 | 0.127 | 2.19 | 26.66 | 0.735 | 0.109 | 6.01 | 29.95 | 0.849 | 0.088 | 3.05 | 24.87 | 0.875 | 0.093 | 2.81 |
| D-Flow | 26.77 | 0.650 | 0.162 | 25.04 | 28.59 | 0.760 | 0.174 | 155.85 | 25.49 | 0.639 | 0.163 | 155.61 | 32.21 | 0.883 | 0.065 | 158.33 | 26.65 | 0.812 | 0.107 | 71.10 |
| Flow-Priors | 29.61 | 0.755 | 0.166 | 31.44 | 27.26 | 0.725 | 0.188 | 31.47 | 24.00 | 0.579 | 0.268 | 31.46 | 33.00 | 0.910 | 0.051 | 32.00 | 26.23 | 0.810 | 0.128 | 31.43 |
| PnP-GS | 32.08 | 0.872 | 0.121 | **0.04** | 27.21 | 0.747 | 0.347 | 1.99 | 21.42 | 0.570 | 0.420 | 1.71 | 28.74 | 0.839 | 0.158 | 0.80 | – | – | – | – |
| PnP-Diff | 31.00 | 0.839 | 0.189 | 1.26 | 28.20 | 0.745 | 0.324 | 1.27 | 24.16 | 0.646 | 0.389 | 47.09 | 32.30 | 0.886 | 0.058 | 1.35 | – | – | – | – |
| Dual Flow-T | 32.78 | 0.897 | **0.042** | 0.40 | 28.03 | 0.744 | **0.088** | **0.41** | 27.50 | 0.759 | **0.087** | **0.66** | 32.60 | 0.909 | **0.031** | **0.23** | 29.57 | 0.924 | **0.031** | **0.21** |
| Dual Flow-W | **33.33** | **0.907** | 0.065 | 0.44 | **29.59** | **0.803** | 0.203 | 0.42 | **28.21** | 0.789 | 0.118 | 0.72 | 33.99 | 0.931 | 0.040 | **0.23** | **29.74** | **0.928** | 0.035 | 0.22 |

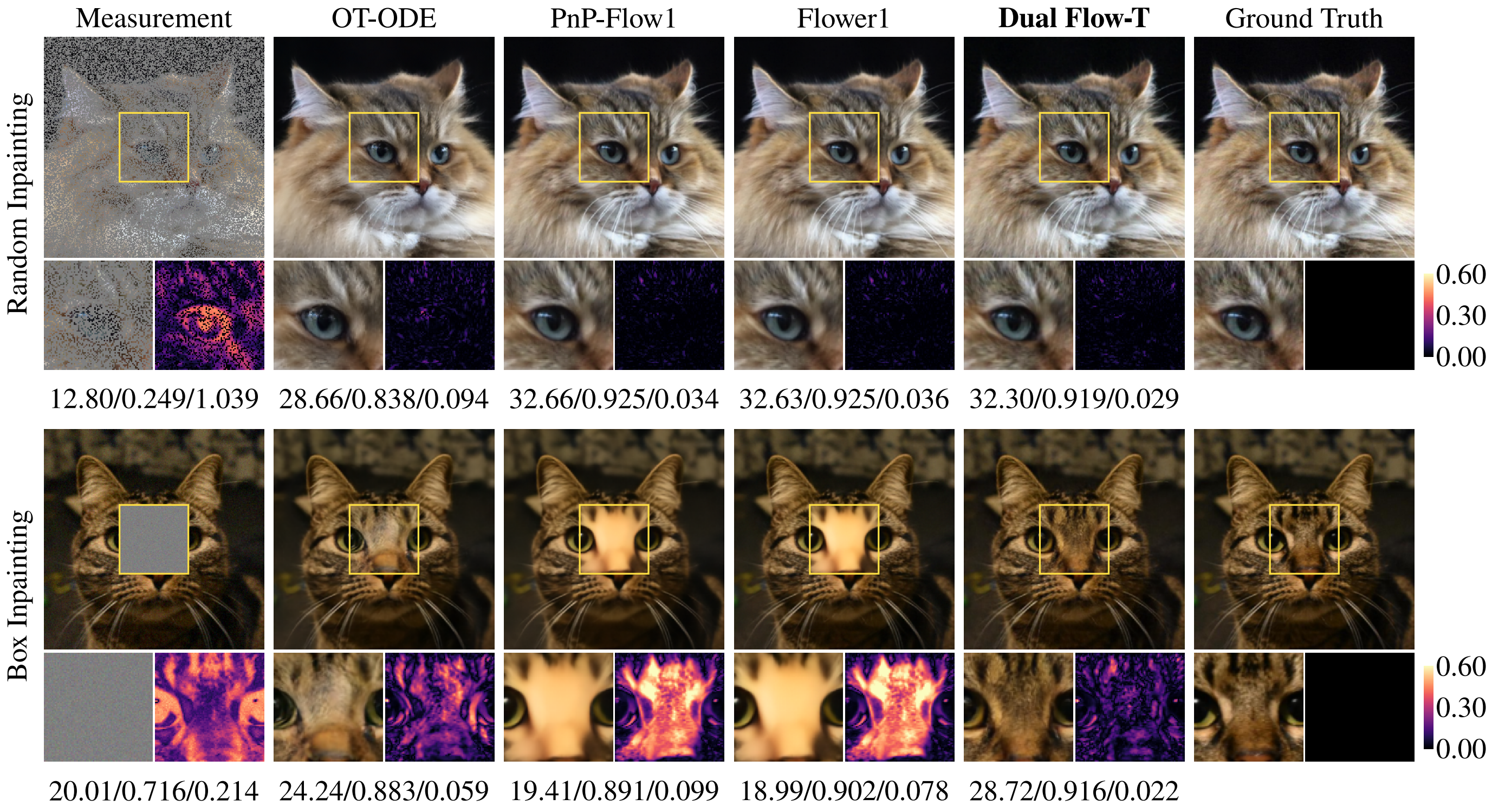

Fig. 9. Qualitative comparison of reconstruction results for random and box inpainting on AFHQ-Cat. Each result shows the full image, a magnified region, and a pixelwise absolute-error map of that region. The values below the images are PSNR, SSIM, and LPIPS, respectively